\documentclass[12pt,nosort]{JHEP3}
\usepackage{exscale,latexsym}
\usepackage[dvips]{graphicx}
\usepackage{axodraw}
\usepackage{cite}

\def\Year{\expandafter\eatPrefix\the\year}
\newcount\hours \newcount\minutes
\def\monthname{\ifcase\month\or
January\or February\or March\or April\or May\or June\or July\or
August\or September\or October\or November\or December\fi}
\def\shortmonthname{\ifcase\month\or
Jan\or Feb\or Mar\or Apr\or May\or Jun\or Jul\or Aug\or Sep\or
Oct\or Nov\or Dec\fi}

\def\TimeStamp{\hours\the\time\divide\hours by60%
\minutes -\the\time\divide\minutes by60\multiply\minutes by60%
\advance\minutes by\the\time%
${\rm \shortmonthname}\cdot   \if\day<10{}0\fi\the\day\cdot
\the\year \qquad\the\hours:\if\minutes<10{}0\fi\the\minutes$}

\newskip\humongous \humongous=0pt plus 1000pt minus 100pt
\def\caja{\mathsurround=0pt}
\def\eqalign#1{\,\vcenter{\openup1\jot \caja
       \ialign{\strut \hfil$\displaystyle{##}$&$
        \displaystyle{{}##}$\hfil\crcr#1\crcr}}\,}
\newif\ifdtup

\newcounter{eqnumber}[section]
\renewcommand{\theeqnumber}{\thesection.\arabic{eqnumber}}
\def\equn{\refstepcounter{eqnumber}
\eqno({\rm \theeqnumber}) }

\def\npb#1#2#3{{\rm Nucl. Phys. B}{\bf  #1}, #3 (#2)}
\def\plb#1#2#3{{\rm Phys. Lett. B}{\bf #1}:#3 (#2)}

\def\prl#1#2#3{{\rm Phys. Rev. Lett. } {\bf  #1}:#3 (#2)}

\def\tree{{\rm tree}}

\newbox\charbox
\newbox\slabox
\def\s#1{{      
        \setbox\charbox=\hbox{$#1$}
        \setbox\slabox=\hbox{$/$}
        \dimen\charbox=\ht\slabox
        \advance\dimen\charbox by -\dp\slabox
        \advance\dimen\charbox by -\ht\charbox
        \advance\dimen\charbox by \dp\charbox
        \divide\dimen\charbox by 2
        \raise-\dimen\charbox\hbox to \wd\charbox{\hss/\hss}
        \llap{$#1$}
}}

\def\spa#1.#2{\left\langle#1\,#2\right\rangle}
\def\spb#1.#2{\left[#1\,#2\right]}
\def\spba#1.#2.#3{\left[#1|#2|#3\right\rangle}
\def\spab#1.#2.#3{\left\langle#1|#2|#3\right]}
\def\spaa#1.#2.#3{\left\langle#1|#2|#3\right\rangle}
\def\spbb#1.#2.#3{\left[#1|#2|#3\right]}
\def\lor#1.#2{\left(#1\,#2\right)}

\catcode`@=11  

\def\Slash#1{\slash\hskip -0.22 cm #1}

\def\eps{\epsilon}

\def\Ord{{\cal O}}

\def\e{\epsilon}

\def\hf{{\textstyle {1\over 2}}}

\def\la{\langle}
\def\ra{\rangle}
\def\oneloop{{1 \mbox{-} \rm loop}}

\def\lsl{\not{\hbox{\kern-2.3pt $\ell$}}}
\def\Psl{\not{\hbox{\kern-2.3pt $P$}}}
\def\ksl{\not{\hbox{\kern-2.3pt $k$}}}
\def\twosl{\not{\hbox{\kern-2.3pt $2$}}}
\def\fivesl{\not{\hbox{\kern-2.3pt $5$}}}
\def\SoftGrav{\mathop{{\cal S}^{\rm gravity}}\nolimits}

\def\rg{r_{\Gamma}}

\def\spa#1.#2{\left\langle#1\,#2\right\rangle}
\def\spb#1.#2{\left[#1\,#2\right]}
\def\lor#1.#2{\left(#1\,#2\right)}

\def\sand#1.#2.#3{%
  \left\langle\smash{#1}{\vphantom1}\right|{#2}%
  \left|\smash{#3}{\vphantom1}\right\rangle}
\def\sandp#1.#2.#3{%
  \left\langle\smash{#1}{\vphantom1}^{-}\right|{#2}%
  \left|\smash{#3}{\vphantom1}^{+}\right\rangle}
\def\sandpp#1.#2.#3{%
  \left\langle\smash{#1}{\vphantom1}^{+}\right|{#2}%
  \left|\smash{#3}{\vphantom1}^{+}\right\rangle}
\def\sandmm#1.#2.#3{%
  \left\langle\smash{#1}{\vphantom1}^{-}\right|{#2}%
  \left|\smash{#3}{\vphantom1}^{-}\right\rangle}
\def\sandpm#1.#2.#3{%
  \left\langle\smash{#1}{\vphantom1}^{+}\right|{#2}%
  \left|\smash{#3}{\vphantom1}^{-}\right\rangle}
\def\sandmp#1.#2.#3{%
  \left\langle\smash{#1}{\vphantom1}^{-}\right|{#2}%
  \left|\smash{#3}{\vphantom1}^{+}\right\rangle}

\def\Aloop{A^{\rm 1\hbox{-}loop}}
\def\Mloop{M^{\rm 1\hbox{-}loop}}

\def\Atree{A^{\rm tree}}

\def\dlips{dLIPS}
\def\NeqEight{{\cal N} = 8}
\def\NeqFour{{\cal N} = 4}

\def\Mtree{M^{\rm tree}}

\def\Li{\mathop{\hbox{\rm Li}}\nolimits}

\def\L{\left(}\def\R{\right)}
\def\dlips{d{\rm LIPS}}

\def\cg{\hat{r}_\Gamma}
\def\Fact{{\cal F}}
\def\Soft{{\cal S}}

\def\BRQ#1#2#3{\la #1|{#2}|#3\ra}
\def\BBRQ#1#2#3{[ #1|{#2}|#3\ra}

\def\small{}

\def\BRi#1#2#3{[ #1|{K}_{#3}|#2\ra}
\def\BRT#1#2#3#4{\la #1|{K}_{#2}K_{#3} |#4\ra}

\def\t(#1,#2,#3){t_{#1#2#3}}
\def\s(#1,#2){s_{#1#2}}
\def\spaN(#1,#2){\left\langle#1#2\right\rangle}
\def\spbN(#1,#2){\left[#1#2\right]}
\def\spabt(#1,\{ #2,#3,#4\},#5){\langle#1\,|P_{#2 #3 #4}|#5]}
\def\spabtt(#1,\{#2,#3,#4\},#5){\langle#1|P_{#2 #3 #4}|#5]}
\def\spbatt(#1,\{#2,#3,#4\},#5){[#1|P\!_{#2 #3 #4}|#5\rangle}
\def\spbattN(#1,\{#2,#3\},#4){[#1|P\!_{#2 #3}|#4\rangle}
\def\spbattt(#1,\{#2,#3\},#4){[#1|P\!_{#2 #3}|#4\rangle}
\def\spaattt(#1,\{#2,#3\},\{#4,#5\},#6){\langle#1|P\!_{#2 #3}P\!_{#4 #5}|#6\rangle}
\def\spabt(#1,\{#2,#3,#4\},#5){\langle#1|P_{#2 #3 #4}|#5]}
\def\spabtt(#1,\{#2,#3,#4\},#5){\langle#1|P_{#2 #3 #4}|#5]}
\def\spbatt(#1,\{#2,#3,#4\},#5){[#1\,|P_{#2 #3 #4}|#5\rangle}

\newcommand{\kspa}[2]{\langle #1#2\rangle}
\newcommand{\kspb}[2]{[#1#2]}
\newcommand{\kspab}[3]{\langle #1|K_{#2}|#3]}
\newcommand{\kspba}[3]{[#1|K_{#2}|#3\rangle}
\newcommand{\kspaa}[4]{\langle #1|K_{#2}K_{#3}|#4\rangle}
\newcommand{\kinv}[1]{t_{#1}}

\title{The No-Triangle Hypothesis for $\NeqEight$ Supergravity}

\author{N.~E.~J.~Bjerrum-Bohr${}^{1,2}$, David~C.~Dunbar${}^1$,
Harald Ita${}^1$,
Warren~B.~Perkins${}^1$ and Kasper Risager${}^3$
\\
${}^1$
Department of Physics,
Swansea University,  Swansea, SA2 8PP, UK
\\
${}^2$
Institute for Advanced Study, Princeton,
NJ 08540, USA
\\
${}^3$
Niels Bohr Institute,
Blegdamsvej 17, DK-2100, 
Copenhagen,
Denmark}

\preprint{hep-th/0610043\\ SWAT-06-473}

\abstract{We study the perturbative expansion of $\NeqEight$
supergravity in four dimensions from the viewpoint of the
``no-triangle'' hypothesis, which states that one-loop graviton
amplitudes in $\NeqEight$ supergravity only contain scalar box
integral functions. Our computations constitute a direct proof at
six-points and support the no-triangle conjecture for seven-point
amplitudes and beyond.}


\begin{document}
\section{Introduction}

$\NeqEight$ supergravity~\cite{NequalEight} is a remarkable theory,
being the maximally supersymmetric field theory containing gravity
that is consistent with unitarity. It is a beautiful but complicated
theory containing (massless) particles of all spins ($\leq 2$) whose
interactions are constrained by a large symmetry group.

This article explores the perturbative expansion of this theory. It
has been postulated that the perturbative expansion of this theory
is more akin to that of \linebreak $\NeqFour$ super Yang-Mills theory than
expected from its known symmetries. In particular, it is
hypothesised that the one-loop amplitudes can be expressed as scalar
box functions with rational coefficients~\cite{BeBbDu}. We provide
considerable evidence for this ``no-triangle hypothesis'' by
examining the behaviour of physical on-shell amplitudes.

This dramatic simplification of the one-loop amplitudes is
presumably a signature of an undiscovered symmetry or principle
present in $\NeqEight$ supergravity. These simplifications do not
occur on a ``diagram by diagram'' basis in any current expansion
scheme, instead they arise only when the diagrams are summed.
Theories of supergravity in four dimensions are one (and two) loop
finite~\cite{NeightCounter}. Since the box functions are UV finite,
the simplifications we see are certainly consistent with these
arguments. However the cancellations are considerably stronger than they demand: 
for example theories with ${\cal N} <8$
supergravity are UV finite at one-loop but the one-loop amplitudes are not
merely box functions.

In this article, we consider one-loop amplitudes and care must be used
in extending the implications beyond one-loop.  However, we do expect
the higher loops to have a softer UV structure than previously
thought\cite{BDDPR}. This opens the door to the possibility that
$\NeqEight$ supergravity may, like $\NeqFour$ super Yang-Mills, be a
finite theory in four dimensions.

\section{The No-triangle Hypothesis}

\subsection{Background: One-loop Amplitudes}

First we review the general structure of one-loop amplitudes in
theories of massless particles. Consider the general form for an
$n$-point amplitude obtained from, for example, a Feynman diagram
calculation~\footnote{For simplicity we restrict ourselves to
covariant gauges with Feynman gauge-like propagators $\sim 1/p^2$},
$$
\Mloop_n(1,\cdots,n)= \sum_{\rm Feynman\,\, diagrams}
I_{r}[ P^{m}(l,\{k_i,\eps_i\}) ]\,,
\equn
$$
where each $I_r$ is a loop momentum integral with $r$
propagators in the loop and numerator $P^m(l,\{k_i,\eps_i\})$. Here
$k_i$ denotes the external (massless) momenta, $\eps_i$ denotes the
polarisation tensors of the external states and $l$ denotes the loop
momentum. For clarity we suppress the $k_i$ and $\eps_i$ labels. In
general the numerator is a polynomial of degree $m$ in the loop
momentum. The value of $m$ depends on the theory under
consideration.
The summation is over all possible diagrams. 

We choose to organise
the diagrams according to the number of propagators in the loop, $r$ . For
$r=n$ the integral will have only massless legs, while for $r<n$ at
least one of the legs attached to the loop will have momentum,
$K=k_a+\cdots k_b$, which is not null, $K^2 \neq 0$. We will call
these massive legs (although it is a slight misnomer in a purely
massless theory).

An important technique for dealing with these integrals is that of
Passarino-Veltman Reduction~\cite{PassVelt}
which reduces any $r$-point integral to a sum of $(r-1)$ point
integrals ($r>4$),
$$
I_r[ P^m(l) ] \longrightarrow \sum_i \,I_{r-1}^i [P^{m-1}(l)]\,.
\equn
$$
We will be evaluating the loop momentum integrals by dimensional regularisation
in $D=4-2\eps$ and working to ${\cal O}(\eps)$. 
In the reduction the degree of the loop momentum polynomial is also
reduced by $1$ from $m$ to $(m-1)$. The $(r-1)$ point functions
appearing are those which may be obtained from $I_r$ by contracting
one of the loop legs. 
This process can be iterated until we obtain four point integrals,
$$
I_r[ P^m(l) ] \longrightarrow \sum_i \,I_{4}^i [P^{m-(r-4)}(l)]\,.
\equn\label{PassVeltFour}
$$
The four point integrals reduce 
$$
 I_4^i[ P^{m'}(l) ] \longrightarrow c_i\, I_4^i[1] +
 \sum_{j}\,  I_3^{j} [P^{m'-1}(l)]\,,
\equn
$$
where we now have the ``scalar box functions'', $I_4[1]$, whose loop
momentum polynomials are just unity. The coefficients $c_i$ are
rational functions of the momentum invariants of the amplitude (By
rational we really mean non-logarithmic, since these coefficients
may contain Gram determinants.) Similarly, reduction of polynomial
triangles gives scalar triangles plus tensor bubble integral
functions,
$$ I_3^j[ P^m(l) ] \longrightarrow   d_j\, I_3^j [1] + \sum_{k}\,
 I_{2}^{k} [P^{m-1}(l)]\,.
\equn
$$
Finally we can express the tensor bubbles as scalar bubble functions
plus rational terms,
$$
 I_{2}^k [P^{m}(l)] =e_k \, I_2^k [1] + R +O(\eps)
\equn
$$
Consequently any one-loop amplitude can be reduced to the form,
$$
 \Mloop_n(1,\cdots,n)=\sum_{i\in \cal C}\, c_i\, I_4^{i}
 +\sum_{j\in \cal D}\, d_{j}\, I_3^{j}
 +\sum_{k\in \cal E}\, e_{k} \,   I_2^{k}
+R\, +O(\eps) ,
\equn
$$
where the amplitude has been split into a sum of integral functions
with rational coefficients and a rational part. The sums run over
bases of box, triangle and bubble integral functions: ${\cal
C},\,{\cal D}$ and ${\cal E}$. Which integral functions appear in a
specific case will depend on the theory and process under
consideration, as will be discussed below.

\subsection{$\NeqFour$ Super Yang-Mills Amplitudes}
For Yang-Mills amplitudes the three-point vertex is linear in
momentum, so generically an $r$-point integral function has a loop
momentum polynomial of degree $r$. In general, a Passarino-Veltman
reduction gives one-loop amplitudes containing all possible integral
functions.

For supersymmetric theories cancellations between the different
types of particle circulating in the loop lead to a reduction in
the order of the loop momentum polynomial. For $\NeqFour$ super
Yang-Mills amplitudes, formalisms exist where four powers of loop
momentum cancel and the generic starting point for the reduction is
a polynomial of degree $m=(r-4)$. This implies that the amplitude
consists only of box and higher point integrals which, via a
Passarino-Veltman reduction~(\ref{PassVeltFour}), give a 
very restricted set of
functions: namely scalar box-functions,
$$
\Aloop_{\NeqFour}\, = \, \sum_{i\in \cal C}\, c_i\, I_4^{i}\,.
\equn
$$

These cancellations can be more or less transparent depending on
the gauge fixing and computational scheme employed. In general the manifest
diagram by diagram cancellation is less than the maximal four
powers. Schemes in which these cancellations are manifest include
the Bern-Kosower string based rules~\cite{StringBased} (where
technically the cancellation occurs at the level of Feynman
parameter polynomials) and well chosen background field gauge
schemes~\cite{Mapping}. In less favourable schemes cancellations
between diagrams occur relatively late in the calculation.

\subsection{$\NeqEight$ Supergravity Amplitudes}
Computation schemes for gravity calculations tend to be rather more
complicated than for Yang-Mills as the three-point vertex is
quadratic in momenta and so the loop momentum polynomial is of
degree $2r$~\cite{DeWitt}. For maximal supergravity we expect to see
considerable cancellations.

In string theory, closed strings contain gravity and open strings
contain gauge theories, so the heuristic relation,
$$
\hbox{closed string} \sim \hbox{(open-string)} \times
                          \hbox{(open-string)} \,,
\equn
$$
suggests a relationship between amplitudes of the form,
$$
\hbox{gravity} \sim \hbox{(Yang-Mills)} \times
                          \hbox{(Yang-Mills)} \,,
\equn
$$
in the low energy limit. For tree amplitudes this relationship is
exhibited by the Kawai-Lewellen-Tye relations~\cite{KLT}. Even in
low energy effective field theories for
gravity~\cite{Donoghue:1994dn} the KLT-relations can be seen to link
effective operators~\cite{EffKLT}. The KLT-relations also hold
regardless of massless matter content~\cite{BDWGravity}. For
one-loop amplitudes we expect such relations for the {\it
integrands} of one-loop amplitudes rather than the amplitudes
themselves. Indeed, the equivalent of the Bern-Kosower rules for
gravity~\cite{StringGravity,DunNor} give an initial loop momentum
polynomial of degree,
$$
2r-8 =2(r-4)\,.
\equn
$$
This power counting is consistent with the heuristic expectation
of string theory.

Using this power counting, reduction for $r>4$ leads to a sum of tensor
box integrals with integrands of degree $r-4$ which would then reduce
to scalar boxes {\it and} triangle, bubble and rational functions,
\vspace{-0.1cm}
$$
\Mloop_{\NeqEight}\,=\, \sum_{i\in \cal C}\, c_i\, I_4^i +\sum_{j\in \cal D}\, d_j\, I^j_3 +\sum_{k\in\,\cal E}\, e_k\, I^k_2 +R\,,
\equn
$$ where we expect that the triangle functions $I_3$ are present for
$n\geq 5$, the bubble functions $I_2$ for $n \geq 6$ and the
rational terms for $n\geq 7$. Note that functions, other than the
scalar boxes, only appear after reduction.

\subsection{The No-triangle Hypothesis}
The ``No-triangle hypothesis'' states that any one-loop amplitude of
$\NeqEight$ supergravity is a sum of box integral functions
multiplied by rational coefficients,
$$
\Mloop_{\NeqEight}= \sum_{i\in \cal C}\, c_i\, I_4^i\,.
\equn
$$
The hypothesis originates from explicit computations which show that
despite the previous power counting arguments, one-loop amplitudes
for $\NeqEight$ supergravity have a form analogous to that of
one-loop $\NeqFour$ super Yang-Mills amplitudes.

The first definite calculation of a one-loop amplitude for both
$\NeqFour$ super Yang-Mills and $\NeqEight$ supergravity was
performed by Green, Schwarz and Brink~\cite{GSB}. By taking the low
energy limit of string theory, they obtained the four point one-loop
amplitudes:
$$
\eqalign{
\Aloop_{(1,2,3,4)}
&=st \times \Atree_{(1,2,3,4)} \times I_4(s,t)\,,
\cr
\Mloop_{(1,2,3,4)}
&=stu\times \Mtree_{(1,2,3,4)}%
\Bigl( I_4(s,t)+I_4(s,u)+I_4(t,u)
\Bigr)\,.
\cr}
\equn
$$
Here $I_4(s,t)$ denotes the scalar box integral with attached legs
in the order $1234$ and $s$, $t$ and $u$ are the usual Mandelstam
variables. The above Yang-Mills amplitude is the leading in colour
contribution. For gravity amplitudes we suppress factors of $\kappa$
( $\kappa^{n-2}$ for tree amplitudes and $\kappa^n$ for one-loop
amplitudes.)  Although only composed of boxes, this gravity
amplitude is consistent with the power counting of $2(r-4)$ with
$r\leq n$.

Beyond four-point we expect to find contributions from other
integral functions in addition to the boxes.  However in
ref.~\cite{multileg} the five and six-point {\it MHV}~\footnote{
Amplitudes are conveniently organised according to the number of
negative helicity external states.  For amplitudes with
``all-positive'' or ``one negative the remaining positive'' helicity
configurations the tree amplitudes vanish for any gravity theory and
the loop amplitudes vanish for any supergravity theory. The first
non-vanishing amplitudes are those with two negative helicity
gravitons, known as ``Maximally Helicity Violating'' or
{\it MHV} amplitudes. Amplitudes with three negative helicity gravitons
are ``next-to-{\it MHV}'' or {\it NMHV} amplitudes. Amplitudes with
exactly two positive helicity gravitons and the remaining negative
helicity can be obtained by conjugation and are known alternatively
as ``googly amplitudes'' or, as used by us, $\overline{\it MHV}$.}
amplitudes were evaluated using unitarity techniques and shown to
consist solely of box integral functions. It was conjectured that
this behaviour continued to all {\it MHV} amplitudes and an all-$n$
ansatz consisting of box functions was presented. This ansatz was
also consistent with factorisation.  In ref.~\cite{BeBbDu} is was
postulated that this was a general feature of $\NeqEight$
amplitudes. In ref.~\cite{BDIgravity} the hypothesis was explored
for the six-point {\it NMHV} amplitude and it was shown that the boxes
alone gave the correct IR behaviour of the amplitude.

In this paper we aim to present further evidence in favour of the
``no-triangle hypothesis''. While we fall short of presenting a
proof, we feel that the weight of evidence is compelling. The
evidence is based on IR structure, unitarity and factorisation. In
the six-point case this evidence does constitute a proof.

\section{Evidence For The No-triangle Hypothesis}

We use a range of techniques to study different parts of the
amplitude: unitarity, factorisation and the singularity structure of
the on-shell physical amplitudes. Our arguments are complete for $n
\leq 6$ point amplitudes. Fortunately there has been considerable
progress in computing one-loop amplitudes inspired by the duality
with twistor space~\cite{BDDK7,BDKn,loopA,loopB,loopC,loopD,
loopE,loopF,loopG,loopH,loopI,Brandhuber:2004yw,
BrittoN=1,BBDI2005}: we will freely use many of these new
techniques.

We use arguments based on the IR divergences of the amplitude to
conclude that the one and two-mass triangles must vanish. We use a
study of the two-particle cuts to deduce that the bubble integrals
are absent and, by numerically examining triple cuts, we show that
the coefficients of three-mass triangles vanish. Finally we use
factorisation arguments to discuss the rational pieces of the
amplitude.

\subsection{IR: Soft Divergences}
The expected soft divergence of an $n$-point one-loop
graviton amplitude~\cite{Dunbar:1995ed} is,
$$
\eqalign{
M^{\rm one-loop}_{(1,2,\ldots, n)} \Bigl|_{\rm soft}
=
{i\kappa^2 \over (4\pi)^2 }
\bigg[
{\sum_{i<j}  s_{ij} \ln[-s_{ij} ]
\over 2\epsilon}
\bigg]
\!\!\times\! M^{\rm tree}_{(1,2,\ldots, n)}\,.
\cr}
\label{IRamplitudeEQ}
\equn
$$
(The factors of $\kappa$ have been reinstated in the amplitudes
within this equation.) For a general amplitude the boxes with three
or fewer massive legs, the one and two mass triangles and the bubble
integrals all have $1/\eps$ singularities which can contribute to
the above.

A necessary condition for the no-triangle hypothesis is that the box
contributions alone yield the complete $1/\eps$ structure. In other
words,
$$
 \sum_{i\in \cal C} \,c_i\, I_4^i {\Big|}_{1/\eps}
={ i\over (4\pi)^2}  \bigg[{\sum_{i<j}  s_{ij} \ln[-s_{ij} ]
\over 2\epsilon}
\bigg]
\!\!\times\! M^{\rm tree}_{(1,2,\ldots, n)}\,.
\equn
$$

If this condition is satisfied, it implies the vanishing of a large
number of the triangle coefficients, specifically that the one and
two-mass triangle functions are not present. The one- and two-mass
triangles are actually not an independent set of integral functions.
As shown in the appendix they can be replaced by a set of basis
functions,
$$
G(-K^2)= {(-K^2)^{-\eps}  \over\e^2} \;
 = { 1 \over \eps^2 } - { \ln ( -K^2 ) \over \eps } +{\rm finite}\,,
\equn
$$
where the set of $G$'s runs over all the independent momentum
invariants, $K^2$, of the amplitude. These functions plus the boxes
then give the only $\ln( -K^2)/\eps$ contributions to the amplitude
since the $1/\eps$ terms in bubbles do not contain logarithms. If
the boxes completely reproduce the required singularity, the
coefficients of the $G$ functions must be zero and consequently the
coefficients of the one- and two-mass triangles can be set to zero,
$$
d_{1m,i}=d_{2m,i} =0\,.
\equn
$$
Having the correct soft behaviour only imposes a single constraint
on the sum of the bubble coefficients,
$$
\sum_i e_i=0\,,
\equn
$$
and, importantly, places no constraint on the three-mass triangles
as they are IR finite.

To verify the IR behaviour, one must know the box coefficients.
Fortunately, there has been considerable progress in computing the
box coefficients in gauge theory.  Box coefficients may be
determined using unitarity~\cite{BDDKa,BDDKb}. In
ref.~\cite{BrittoUnitarity}, Britto, Cachazo and Feng showed that
quadruple cuts can be used to algebraically obtain box coefficients
from the four tree amplitudes at the corners of the cut box.
Specifically, if we consider an amplitude containing a scalar box
integral function, the coefficient of this function is given by the
product of four tree amplitudes with on-shell cut
legs~\cite{BrittoUnitarity},
$$
\eqalign{
 c={ 1 \over 2 } \sum_{h_i\in\,\cal S}&
  \biggl( \Mtree\big( (-\ell_1)^{-h_1},i_1,  \ldots,i_2,(\ell_2)^{h_2}\big) \times
  \Mtree\big( (-\ell_2)^{-h_2},i_3,\ldots,i_4,(\ell_3)^{h_3}\big)\cr
&\hspace{-0.5cm}\times \Mtree\big( (-\ell_3)^{-h_3},i_5,\ldots,i_6,(\ell_4)^{h_4}\big) \times
 \Mtree\big( (-\ell_4)^{-h_4},i_7,\ldots,i_8,(\ell_1)^{h_1}\big) \biggr)\,.
\cr}\hspace{-1cm} \equn
$$
Here ${\cal S}$ indicates the set of possible particle and helicity
configurations of the legs $\ell_i$ which give a non-vanishing product
of tree amplitudes We often denote the above coefficient by the clustering on the legs, 
$c^{[\{i_1\cdots i_2\},\{i_3\cdots i_4\},\{i_5\cdots i_6\},\{i_7\cdots i_8\}]}$.
In the above the tree amplitudes at massless
corners require analytic continuation.

The box coefficients may also be obtained from the known box
coefficients for $\NeqFour$ Yang-Mills~\cite{BDDKb,BDDK7,BDKn} by
squaring and summing~\cite{BeBbDu}. For example for the three-mass
boxes within the seven-point {\it NMHV} amplitude we have,
$$
\eqalign{
 c_{\NeqEight}^{[1^-,\{4^+5^+\},\{2^-3^-\},\{6^+7^+\}]}
&=
2 s_{23}s_{45}s_{67} \; c_{\NeqFour}^{[1^-,\{4^+5^+\},\{2^-3^-\},\{6^+7^+\}]}
                     \; c_{\NeqFour}^{[1^-,\{5^+4^+\},\{3^-2^-\},\{7^+6^+\}]}
\cr},
\equn
$$
which allows us to obtain the $\NeqEight$ coefficients from 
the $\NeqFour$ box coefficients.

We have computed the IR behaviour of the six and seven-point {\it NMHV}
amplitudes. The six-point box coefficients are given in
ref.~\cite{BDIgravity} and the seven-point box coefficients are
given in appendix~\ref{SevenPointAppendix}. In both cases amplitudes were
constructed using these box-coefficients and, after some computer algebra,  
the resultant amplitudes were found to 
reproduce the complete IR behaviour. This  allows us to conclude
that,
$$
d_{2m,i}=d_{1m,i} =0  \hbox{ for } n=6,7.
\equn
$$

\subsection{Two-Particle Cuts}
A general unitarity cut of the amplitude $M_n(1,2,\ldots n)$ in the
channel carrying momentum $P=k_i+\ldots k_j$, is given by a sum of
phase space integrals of products of tree amplitudes,
$$
\eqalign{C_{i\cdots j}
=i\sum_{h_1,h_2\in\,\cal S'}\,\int \dlips (-l_1,l_2)\;
&\Mtree\big( (-l_1)^{-h_1},i,\cdots, j,(l_2)^{h_2}\big)\cr
&\hspace{1cm}\times\Mtree\big( (-l_2)^{-h_2},j+1,\cdots, i-1,(l_1)^{h_1}\big)\,,
}\hspace{-1cm}\equn\label{cut}
$$
where ${\cal S'}$ denotes the helicities of the particles from the
$\NeqEight$-multiplet that can run in the loop. This unitarity cut
is equal to the leading discontinuity of the loop amplitude,
$$
\hspace{-4.8cm}
\eqalign{&\sum_{i\in \cal C}\, c_i\, I_4^i +\sum_{j\in\,\cal D}\, d_j\, I^j_3 +\sum_{k\in\,\cal E}\, e_k\, I^k_2\,{\Big|}_{\rm Disc}
\cr&=
i\,\int \dlips (-l_1,l_2)
\left[ \sum_{i\in \cal C'} { c_{i} \over (l_1-K_{i,4})^2 (l_2-K_{i,2})^2 }
+\sum_{j\in\,\cal D'} {d_{j} \over (l_1-K_{j,3})^2 }
+e_{k'}
\right]\,.
}
\hspace{-4cm}\equn\label{discontinuities}
$$
The sets of box and triangle functions that contribute to a given cut
are denoted by ${\cal C'}$ and ${\cal D'}$ respectively and the
single bubble function that contributes is labelled by $k'$. In
principle the coefficients of all the integral functions can be
obtained by performing all possible two-particle cuts. In practice
it is often simpler to determine the box and triangle coefficients
by other means before using the two-particle cuts to determine the
bubble terms. The rational pieces of the amplitude are not
``cut-constructible''~\cite{BDDKa,BDDKb}.

To show that a given integral function is absent from the amplitude
we have to show that its contribution to the cut integral vanishes.
This test may be done by either evaluating the cut integral
explicitly or, equivalently, by algebraically reducing the integrand to
a sum of constant coefficients times specific products of
propagators, that are the {\it signatures} of the cuts of 
specific integral functions.

\subsection{Bubble Integrals from the two-particle cuts}

In this section we will show, by explicit computation of the
two-particle cuts, that all bubble integrals in the
six-point amplitudes vanish. 
These arguments can also be used to show
that bubble integrals are absent from all the cuts of all-n one-loop
{\it MHV} amplitudes as discussed in section~(\ref{usingthelargeztest}).

Recently, the realisation that Yang-Mills amplitudes are dual to a
twistor string theory~\cite{WittenTopologicalString} has given
considerable impetus to gauge theory calculations. In particular, it
appears that the two-particle cuts can be efficiently calculated if
expressed in spinor or twistor variables~\cite{BrittoN=1}.

Consider the two-particle cut,
$$
\eqalign{
C_{12}\!=\!&
i \int d\mu\,
M^{\rm tree}_4
\big((-l_1)^+,1^-,2^-,(l_2)^+\big)\times
M^{\rm tree}_6
\big( (l_1)^-,3^-,4^+,5^+,6^+,(-l_2)^-\big)\,,\cr
&{\hspace{-.3cm}\hbox{ where } } d\mu=d^4l_1d^4l_2
\delta^{(+)}(l_1^2)
\delta^{(+)}(l_2^2)
\delta^{(4)} (l_1-l_2-k_1-k_2) ,
\cr}
\hspace{-1cm}\equn\label{cutintA}
$$
with a graviton running in the loop. We denote the integrand by
${\cal I }(l_1,l_2)$. Setting $l_1=t\ell$ with $\ell_{a\dot
a}=\lambda_a \tilde\lambda_{\dot a}$, the measure
becomes~\cite{Cachazo:2004kj,BrittoN=1},
$$
d^4l_1\delta^{(+)}(l_1^2)=tdt\spa{\lambda}.{d\lambda}\spb{\tilde\lambda}.{d\tilde\lambda}\,,
\equn
$$
so that the cut becomes,
$$
\hspace{-5.3cm}\eqalign{
C_{12}&=i\!\int\! d\mu \,{\cal I}(l_1,l_2)=i\int_0^\infty\!\!\! tdt\int
\!\spa{\lambda}.{d\lambda}\spb{\tilde\lambda}.{d\tilde\lambda}
\delta^{(+)}\big(P^2-t P_{a\dot a}\lambda^a\tilde\lambda^{\dot a}\big)\,  {\cal I}\big(t\ell,-P-t\ell\big)\cr
&= i\!\int \spa{\lambda}.{d\lambda}\spb{\tilde\lambda}.{d\tilde\lambda}
\frac{P^2}{(P_{a\dot a}\lambda^a\tilde\lambda^{\dot a})^2}
\,{\cal I}\biggr({P^2\over P_{a\dot a}\lambda^a\tilde\lambda^{\dot a}}
      \ell, -P-{P^2\over P_{a\dot a}\lambda^a\tilde\lambda^{\dot
          a}}\ell
\biggr)\,,}
\hspace{-5cm}\equn\label{cutintB}
$$
where $P$ denotes the total momentum on one side of the cut. In the
example above, $P=k_1+k_2$. Powers of $l_1$ within ${\cal I}$ give
rise to powers of $t$ which in turn give rise to extra powers of
$P^2/(P_{a\dot a}\lambda^a\tilde\lambda^{\dot a})$ due to the
$\delta(P^2-t P_{a\dot a}\lambda^a\tilde\lambda^{\dot a})$-function.
Thus in general the cut will be a sum of terms with different powers
of $(P_{a\dot a}\lambda^a\tilde\lambda^{\dot a})$,
$$
C_{12} = \int \spa{\lambda}.{d\lambda}\spb{\tilde\lambda}.{d\tilde\lambda}
\sum_n  { f_n( \lambda,\tilde\lambda) \over (P_{a\dot a}\lambda^a\tilde\lambda^{\dot a})^n }.
\equn
$$
The key observation of~\cite{BrittoN=1,Britto:2006sj} is that the
different classes of integral function that contribute to the cut
can be recognised by the powers of $(P_{a\dot
a}\lambda^a\tilde\lambda^{\dot a})$ that are present. Generically,
any term containing, $1/(P_{a\dot a}\lambda^a\tilde\lambda^{\dot
a})^n$ with $n< 2$ in $C_{12}$ will not generate a contribution
to the coefficient of any bubble integral function. In terms of $t$,
such terms correspond to terms in ${\cal I}$ of the form $t^{m}$
with $m<1$. In the following we show that only terms of this
form arise in two-particle cuts of the six-point one-loop amplitudes
and hence that no bubble integral functions contribute to these
amplitudes.

The {\it NMHV} amplitude $\Mloop(1^-,2^-,3^-,4^+,5^+,6^+)$ has four
inequivalent cuts up to relabelling of external legs; $C_{12}$,
$C_{34}$, $C_{123}$ and $C_{234}$. Of these $C_{12}$ and $C_{123}$
are what we call singlet cuts. These cuts vanish unless the two
outgoing cut legs have the same helicity, implying that these states
can only be gravitons. These singlet cuts are thus independent of
the matter content of the theory and the absence of bubble
functions is independent of the number of supersymmetries. For the
non-singlet cuts, the two outgoing cut legs have opposite helicity
and so the full $\NeqEight$ multiplet contributes. For these cuts,
bubble functions are only absent from the $\NeqEight$ amplitudes.

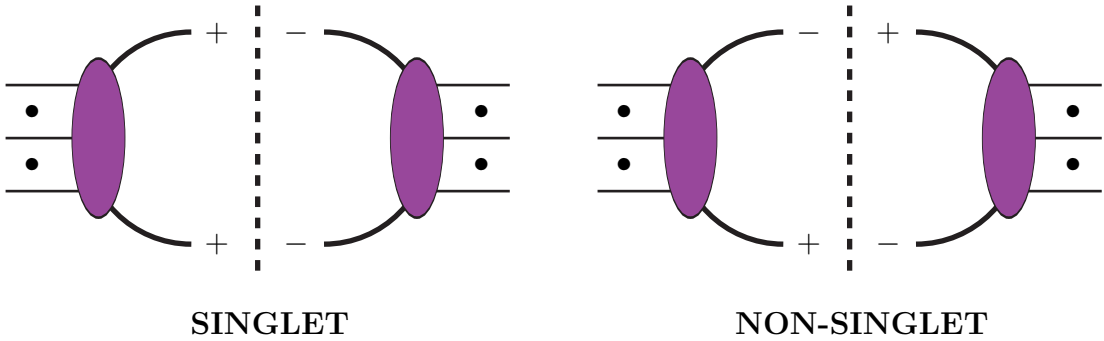
\begin{center}
\begin{picture}(220,100)(0,0)
 \SetWidth{2}
\CArc(70,50)(40,90,270)
\DashLine(95,0)(95,100){4}
\CArc(120,50)(40,270,90)
\Text(80,10)[]{$+$}
\Text(80,90)[]{$+$}
\Text(110,10)[]{$-$}
\Text(110,90)[]{$-$}
\SetWidth{1}
\Line(35,50)(0,50)
\Line(35,70)(0,70)
\Line(35,30)(0,30)
\Text(10,60)[]{$\bullet$}
\Text(10,40)[]{$\bullet$}
\Line(155,50)(190,50)
\Line(155,70)(190,70)
\Line(155,30)(190,30)
\Text(180,60)[]{$\bullet$}
\Text(180,40)[]{$\bullet$}
\COval(35,50)(30,10)(0){Black}{Purple}
\COval(155,50)(30,10)(0){Black}{Purple}
\Text(100,-20)[]{\bf SINGLET}
\end{picture}
\begin{picture}(200,100)(0,0)
 \SetWidth{2}
\CArc(70,50)(40,90,270)
\DashLine(95,0)(95,100){4}
\CArc(120,50)(40,270,90)
\Text(80,10)[]{$+$}
\Text(80,90)[]{$-$}
\Text(110,10)[]{$-$}
\Text(110,90)[]{$+$}
\SetWidth{1}
\Line(35,50)(0,50)
\Line(35,70)(0,70)
\Line(35,30)(0,30)
\Text(10,60)[]{$\bullet$}
\Text(10,40)[]{$\bullet$}
\Line(155,50)(190,50)
\Line(155,70)(190,70)
\Line(155,30)(190,30)
\Text(180,60)[]{$\bullet$}
\Text(180,40)[]{$\bullet$}
\COval(35,50)(30,10)(0){Black}{Purple}
\COval(155,50)(30,10)(0){Black}{Purple}
\Text(100,-20)[]{\bf NON-SINGLET}
\end{picture}
\end{center}
\vspace{1.2cm}

We now examine the four distinct cuts in turn. First we consider
$C_{123}$, as this is the simplest: it is a singlet cut and the tree
amplitudes that appear are either {\it MHV} or $\overline{\it MHV}$
amplitudes. Explicitly the product of tree amplitudes is,
$$
\eqalign{
\Mtree_{\rm MHV}\big((-l_1)^+,& 1^-,2^-,3^-,(l_2)^+\big)\times\Mtree_{\rm MHV}\big((-l_2)^-,4^+,5^+,6^+,(l_1)^-\big).
\cr
&=-\spb{l_1}.{l_2}^8\frac{\spb3.{l_1}\spa{l_1}.1 \spb1.2 \spa2.3-\spa3.{l_1}\spb{l_1}.1 \spa1.2 \spb2.3}{\spb{l_1}.{l_2}\spb{l_1}.1\spb{l_1}.2\spb{l_1}.3\spb{l_2}.1\spb{l_2}.2\spb{l_2}.3\spb1.2\spb1.3\spb2.3}\cr
&\hspace{0.4cm}\times\spa{l_1}.{l_2}^8\frac{\spa6.{l_1}\spb{l_1}.4 \spa4.5 \spb5.6-\spb6.{l_1}\spa{l_1}.4 \spb4.5 \spa5.6}{\spa{l_1}.{l_2}\spa{l_1}.4\spa{l_1}.5\spa{l_1}.6\spa{l_2}.4\spa{l_2}.5\spa{l_2}.6\spa4.5\spa4.6\spa5.6}\,.
}
\hspace{-1cm}\equn$$
This can be simplified to,
$$
\eqalign{&-\frac{\big(P_{123}^2\big)^{10}}{\spb1.2\spb1.3\spb2.3\spa4.5\spa4.6\spa5.6}\times\cr
&\frac{\big(\spb3.{l_1}\spa{l_1}.1 \spb1.2 \spa2.3-\spa3.{l_1}\spb{l_1}.1 \spa1.2 \spb2.3\big)\big(\spa6.{l_1}\spb{l_1}.4 \spa4.5 \spb5.6-\spb6.{l_1}\spa{l_1}.4 \spb4.5 \spa5.6\big)}{
\spb{l_1}.1\spb{l_1}.2\spb{l_1}.3\spa{l_1}.4\spa{l_1}.5\spa{l_1}.6
\prod_{x=1,2,3}\spab{l_1}.P_{123}.x
\prod_{y=4,5,6} \spba{l_1}.P_{123}.y \
}}\equn
$$
Substituting $l_1=t l$ into the above term we find a factor of
$1/t^{4}$ and hence there are no bubble contributions to this cut.

Next we consider $C_{234}$. Again the tree amplitudes are either
{\it MHV} or $\overline{\it MHV}$ amplitudes, but this is a non-singlet cut,
so we must include a summation over the super-multiplet.
{\it MHV}($\overline{\it MHV}$) tree amplitudes with a single pair of
non-graviton particles are related to the corresponding pure
graviton amplitude by simple factors, $X(h)$. The summed integrand is most
naturally  expressed in terms of tree amplitudes with a scalar
circulating in the loop and a {\it $\rho$-factor}. Using a
superscript $s$ to denote a scalar in the loop, we have,
$$\hspace{-0.25cm}
\eqalign{
&\sum_{h\in S'}
\Mtree\big( (-l_1)^{-h},2^-,3^-,4^+,(l_2)^h\big)\times
\Mtree\big((-l_2)^{-h},5^+,6^+,1^-,(l_1)^{h}\big)
\cr
&\hspace{0.2cm}=
\Mtree\big( (-l_1)^{s},2^-,3^-,4^+,(l_2)^s\big)\times
\Mtree\big((-l_2)^{s},5^+,6^+,1^-,(l_1)^{s}\big)
\sum_{h\in S'} X(h) 
\cr
&\hspace{0.2cm}=
\rho\times\Mtree\big( (-l_1)^{s},2^-,3^-,4^+,(l_2)^s\big)\times
\Mtree\big((-l_2)^{s},5^+,6^+,1^-,(l_1)^{s}\big),
\cr}
\equn\label{nonsingterms}
$$
where,
$$
\rho = \sum_{h\in S'} X(h)
\!\!
=\sum_{a=-4}^{a=4} {8!\over (4-a)!(4+a)!}
\left( { x \over y} \right)^{a}\!\! =  { (x+y)^8\over x^4 y^4 }=\frac{\spab1.P_{234}.4 ^8}{\big(\spb4.{l_1} \spb4.{l_2} \spa1.{l_1}\spa1.{l_2}\big)^4}\, .
\equn
$$
The factor $n_h={8!/( (4-a)!(4+a)!)}$ is the multiplicity within the
$\NeqEight$ multiplet of the states of helicity $h=a/2$. 
Rewriting the amplitude in terms of $l_1$ we can count the powers of $t$.
Overall the leading contributions are $O(t^{-4})$, just as in
the singlet case. Once again the cut receives no contributions from
bubble functions.

The remaining cuts are algebraically more complicated, but they
repeat the patterns seen above. $C_{12}$ is a singlet cut involving
the product of a four-point {\it MHV} amplitude, $\Mtree\big(
(-l_1)^+,1^-,2^-,(l_2)^+\big)$, and a six-point {\it NMHV}
amplitude,\linebreak $\Mtree\big((-l_2)^-,3^-,4^+,5^+,6^+,(l_1)^-\big)$. The
six-point {\it NMHV} tree amplitude has only recently been calculated
using on-shell recursion~\cite{Britto:new,BBSTgravity,CSgravity}.
An
explicit form for this amplitude as a sum of fourteen terms is given in
appendix~\ref{sixpointmultiplet}.\footnote{In general much less is
known about gravity tree amplitudes than Yang-Mills amplitudes. Traditional
Feynman diagram approaches tend to be excessively complicated as
evidenced by the computation by Sannon~\cite{DeWitt} of the
four-point tree amplitude. The KLT relations, which express the
gravity amplitudes as sums of permutations of products of two
Yang-Mills amplitudes~\cite{KLT}, are an extremely useful
technique, however the factorisation structure is rather obscure and
the permutation sum grows quickly with the number of legs. Of the
new techniques, the BCF recursion readily extends to gravity
amplitudes~\cite{BBSTgravity,CSgravity} giving useful compact
results. The {\it MHV}-vertex approach of Cachazo, Svr\v cek and Witten
also extends to gravity~\cite{gravCSW} although the correct analytic
continuation of the {\it MHV} gravity vertices is only clear after using
the appropriate factorisation~\cite{Kasper}. Currently, there is no
Lagrangian based proof of these techniques such as exists for
Yang-Mills~\cite{Mansfield}, however we have numerically checked the
expressions for both {\it MHV} vertices and recursion against the KLT
expressions for amplitudes with seven or fewer points.}

We will illustrate here how one of the terms gives a contribution that
vanishes at large $t$. The remaining thirteen terms will follow
analogously and thus we see term-by-term that this cut receives no
contributions from bubble functions. The singlet 12-cut reads,
$$
C_{12}=i\int d\mu\,M_4\big((-l_1)^+,1^-,2^-,(l_2)^+\big)\,
\times\,M_6\big((-l_2)^-,3^-,4^+,5^+,6^+,(l_1)^-\big)\,,
\equn\label{godawfulequation}
$$
where the four-point amplitude is,
$$
M_4\big((-l_1)^+,1^-,2^-,(l_2)^+\big)\,=\,
\frac{i\spa1.2^7\spb1.2}{\spa1.{l_1}\spa1.{l_2}
\spa2.{l_1}\spa2.{l_2}\spa{l_1}.{l_2}^2}\,,
\equn
$$
and the six-point amplitude is given in~\cite{BDIgravity}. We
will in this example analyse the contribution to the cut given by
the term $G_4^{ns}[-l_2,3,5,4,6,l_1]$ in the full amplitude,
$$\eqalign{
M_6\big((-l_2)^-,3^-,4^+,5^+,6^+,(l_1)^-\big)|_{G^{ns}_4[-l_2,3,5,4,6,l_1]}&=\cr &\hspace{-6.5cm}
\frac{i \s(3,5) \s(4,6) \s(l_1,l_2)
\spbatt(5,\{l_2,3,5\},l_1)^7
}{
\spaN(4,6)^2  \spaN(4,l_1)  \spaN(6,l_1)
\spbN(3,5)^2  \spbN(3,l_2)  \spbN(5,l_2) \spbatt(3,\{l_2,3,5\},l_1)
\spbatt(l_2,\{l_2,3,5\},4) \spbatt(l_2,\{l_2,3,5\},6)
\t(3,5,l_2)}\,,}
\equn
$$
so that the integrand of (\ref{godawfulequation}) is,
$$\eqalign{
&\frac{-\spa1\!.2^7\spb1\!.2}{\spa1.{\!l_1}\!\spa1.{\!l_2}\!\spa2.{\!l_1}\!\spa2.{\!l_2}\!\spa{l_1}.{\!l_2}^2}\times
\cr &
\hspace{0.4cm}\frac{ \s(3,5)\s(4,6)\s(l_1,l_2)
\spbatt(5,\{l_2,3,5\},l_1)^7
}{
\spaN(4,6)^2 \spaN(4,l_1) \spaN(6,l_1)
\spbN(3,5)^2 \spbN(3,l_2) \spbN(5,l_2) \spbatt(3\!,\{l_2,3,5\},l_1)
\spbatt(l_2\!,\{l_2,3,5\},4) \spbatt(l_2\!,\{l_2,3,5\},6)
\t(3,5,l_2)
}\,,}\equn
$$
which can be written as,
$${\cal C}\times
\frac{\spbatt(5,\{l_2,3,5\},l_1)^7}{\spa1.{\!l_2}\!\spa1.{\!l_1}\!\spa2.{\!l_2}\!
\spa2.{\!l_1}\!\spa{l_2}.{\!l_1}^2\!
\spaN(4,l_1)\! \spaN(6,l_1)
\spbN(3,l_2)\! \spbN(5,l_2) \spbatt(3,\{l_2,3,5\},l_1)
\spbatt(l_2\!,\{l_2,3,5\},4) \spbatt(l_2\!,\{l_2,3,5\},6)
\t(3,5,l_2)}\,,\equn
$$
where,
 $$\displaystyle{\cal C}=\frac{\s(3,5) \s(4,6) \s(1,2)^2\spa1.2^6}{\spaN(4,6)^2\spbN(3,5)^2}\,.\equn$$\\
Now transforming all $l_2$ into $l_1$ using
$\displaystyle [ X  l_2 ] \rightarrow {[ X | {P_{12} | l_1 \rangle}\over{\spaN(l_2,l_1)}}$ and
$\displaystyle \langle Y  l_2 \rangle \rightarrow {\langle Y | {P_{12} | l_1 ]}\over{\spbN(l_2,l_1)}}$ 
we get,
$$
\frac{s_{12}^2\cal C}{\spa1.2^2}\times
{\cal H}(|l_1\rangle)\times
\frac{1}{\spb2.{l_1}\spb1.{l_1}\t(4,6,l_1)}\,,\equn
$$
where,
$$\displaystyle {\cal H}(|l_1\rangle)=\frac{\spbattN(5,\{4,6\},l_1)^7}{\spaN(1,{l_1})\spaN(2,{l_1})
\spaN(4,l_1) \spaN(6,l_1)
\spbattt(3,\{1,2\},l_1)\spbattt(5,\{1,2\},l_1) \spbattN(3,\{4,6\},l_1)
\spaattt(l_1,\{1,2\},\{4,6\},4) \spaattt(l_1,\{1,2\},\{4,6\},6)}\,.$$
Now we have to count the number of factors of $t$. We get a total count of $1/t^2$ hence no bubbles integral functions are present in the cut.

The remaining $C_{34}$ cut is non-singlet and so we again need to
sum over the multiplet. Explicit forms for the relevant six-point
amplitudes involving an arbitrary pair of particles plus gravitons
are given in appendix~\ref{sixpointmultiplet}. These tree amplitudes
are each a sum of fourteen terms.  As we change the non-graviton
particles, the individually terms in the amplitude each behave like
{\it MHV} amplitudes in that they collect simple multiplicative factors.
Performing the sum over the multiplet term-by-term we find a
$\rho$-factor for each term. Just as in the $C_{234}$ cut, these are
very  important as they introduce large inverse powers of $t$. For
most terms, $\rho \sim 1/t^8$. Again we pick a sample term to
illustrate the process: the other thirteen terms follow analogously.

We will consider the cut,
$$
C_{34}=i\int d\mu\,\sum_{h \in {\cal S}'} M_4\big((-l_1)^h,3^-,4^+,(l_2)^{-h}\big)\,M_6\big((-l_2)^h,5^+,6^+,1^-,2^-,(l_1)^{-h}\big)\,.\equn
$$
The four-point tree amplitude
$M_4\big((-l_1)^h,3^-,4^+,(l_2)^{-h}\big)$ is given by,
$$
\eqalign{
&M_4\big((-l_1)^h,3^-,4^+,(l_2)^{-h}\big)=\frac{i \,\spaN({l_2},3)^7\,\spbN({l_2},3)}{\spaN(3,4)\,\spaN(3,{l_1})\,{\spaN(4,{l_1})}^2\,\spaN({l_2},4)\,\spaN({l_2},{l_1})}
\left(\frac{\spaN({-l_1},3)}{\spaN(l_2,3)}\right)^{4-2h}\,.
}\equn
$$
For the six-point corner we consider a  specific but representative
term from the fourteen in eq~(\ref{6ptmultiplet}),
$$
\eqalign{
&M_6\big((-l_2)^h,5^+,6^+,1^-,2^-,(l_1)^{-h}\big)|_{T_2}\cr&=
\left(-\frac{i \spaN(1,l_2) \spbN(6,l_1)}{\spabtt(6,\{2,6,l_1\},1)}\right)^{4-2h}\times\cr
& \hspace{1.9cm}\frac{-i
   \spaN(2,l_1) \spabtt(1,\{2,6,l_1\},6)^8 \spbN(5,l_2)}{\spaN(1,5) \spaN(1,l_2)
   \spaN(5,l_2) \spabtt(1,\{2,6,l_1\},2) \spabtt(1,\{2,6,l_1\},l_1)
   \spabtt(5,\{2,6,l_1\},6) \spabtt(l_2,\{2,6,l_1\},6) \spbN(2,6) \spbN(2,l_1)
   \spbN(6,l_1) \t(2,6,l_1)}\,.
}\hspace{-3cm}\equn
$$
The particle type dependent factors can be extracted and we find relative to the graviton amplitude,
$$
\eqalign{\rho=&\sum_{h\in\,{\cal S}'} \left(-\frac{i \spaN(1,l_2) \spbN(6,l_1)}{\spbatt(6,\{2,6,l_1\},1)}\frac{\spaN({-l_1},3)}{\spaN(l_2,3)}\right)^{4-2h}\,
=\left(\frac{-\spaN(1,2) \spbN(2,6) \spaN(3,{l_2})+\spaN(1,3)\spba6.P_{34}.{l_2}}{\spbatt(6,\{1,5,-l_2\},1)\spaN({l_2},3)}\right)^8\,.
}\equn
$$
Next we rewrite the cut in terms of the loop momenta $l_2$ using the 
on-shell conditions and $l_1=l_2+k_3+k_4$. The $\rho$ factor already has 
the correct form. The remaining contributions to the cut integral are
then,
$${\cal C}\times
{\cal H}(|l_2\rangle)\times
\frac{\spab2.P_{34}.{l_2} \spbN(5,l_2)}{\spbN(3,{l_2}){\spbN(4,{l_2})}  \spabtt(1,\{1,5,{-l_2}\},2)\spabtt(5,\{1,5,-l_2\},6) \t(1,5,{-l_2})}\,,
\equn$$
where $${\cal C} = \frac{\spbN(3,4)^2}{\spaN(3,4)^2\spaN(1,5)\spbN(2,6)}\,,\equn$$ and
$${\cal H}(|l_2\rangle) = \frac{(-\spaN(1,2) \spbN(2,6) \spaN(3,{l_2})+\spaN(1,3)\spba6.P_{34}.{l_2})^8}
{\spaN({l_2},3)\spaN({l_2},4)\spaN(1,l_2)\spaN(5,l_2)\spba6.P_{34}.{l_2}\spba2.P_{34}.{l_2}\spab{l_2}.P_{15}.6\spaa1.P_{26}P_{34}.{l_2}}\,.\equn$$

We now replace $l_2$ by $l_2=t\, \ell=t\,\lambda\tilde\lambda$ and
do the $t$-integration. With the definitions,
$\spab\ell.{Q_1}.\ell=\spab\ell.{P_{34}}.\ell
\spab1.P_{15}.2-s_{34}\spaN(\ell,1)\spbN(2,\ell)$,
$\spab\ell.{Q_2}.\ell=\spab\ell.{P_{34}}.\ell
\spab5.P_{15}.6-s_{34}\spaN(\ell,5)\spbN(6,\ell)$,
$\spab\ell.{Q_3}.\ell=\spab\ell.{s_{15}P_{34}-s_{34}P_{15}}.\ell$,
we can rewrite the cut as,
$$
\eqalign{
&-{\cal C}\times
s_{34}\times{\cal H}(|l\rangle)\times\spab\ell.P_{34}.\ell \times
\frac{i\spab2.P_{34}.{\ell} \spbN(5,\ell)}
{ \spbN(3,{\ell}){\spbN(4,{\ell})}\spab\ell.Q_1.\ell\spab\ell.Q_2.\ell \spab\ell.Q_3.\ell}
}\,.\equn
$$
It is important to notice that the $\rho$-factor contributes
$\spab\ell.P_{34}.\ell^8$, while the product of the graviton
amplitudes gives rise to $1/\spab\ell.P_{34}.\ell^5$,  with a further factor of
$1/\spab\ell.P_{34}.\ell^2$ coming from the integration measure. The powers
of $\spab\ell.P_{34}.\ell$ are important in that they
indicate the type of integral functions that are present. For the
above term with only single poles in the denominator, bubbles can
only arise from terms carrying a factor of $1/\spab\ell.P_{34}.\ell^2$. We
therefore conclude that no bubbles are present.

By considering all distinct two-particle cuts of the six-point
one-loop {\it NMHV} amplitude we have shown that the amplitude receives
no contributions from bubble integral functions.

\subsection{Triple Cuts}
Having verified that no one or two-mass triangles or bubble integral
functions are present in the amplitude,  we now consider the
three-mass triangle integral function. These have no IR
singularities and so the previous arguments have nothing to say
regarding their absence or presence. In this section we illustrate
how the coefficients of three-mass triangles can be evaluated by
numerically integrating triple cuts of amplitudes. Note that {\it MHV}
amplitudes do not contain triple cuts for any gravity theory so this
is a previously untested class of functions.

Consider a physical triple cut in an amplitude where all three
corners are massive, 
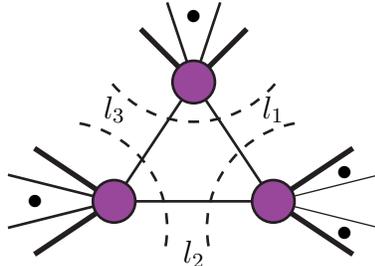
\vspace{-0.3cm}
\begin{center}
\begin{picture}(40,60)(-20,50)
\Line(30,30)(70,40)
\Line(30,30)(70,20)
\SetWidth{2}
\Line(30,30)(60,50)
\Line(30,30)(60,10)
\SetWidth{1}
\Line(30,30)(-30,30)
\Line(-30,30)(0,75)
\Line(30,30)(0,75)
\Line(-30,30)(-70,40)
\Line(-30,30)(-70,20)
\SetWidth{2}
\Line(-30,30)(-60,50)
\Line(-30,30)(-60,10)
\SetWidth{1}
\Text(-60,30)[]{$\bullet$}
\Line(0,75)(-10,105)
\Line(0,75)(10,105)
\SetWidth{2}
\Line(0,75)(-20,95)
\Line(0,75)(20,95)
\Text(0,100)[]{$\bullet$}
\Text(57,41)[]{$\bullet$}
\Text(57,18)[]{$\bullet$}
 \SetWidth{1}
\DashCArc(45,20)(40,100,190){4}
\DashCArc(-50,20)(40,-10,80){4}
\DashCArc(00,100)(40,220,320){4}
\CCirc(30,30){8}{Black}{Purple}
\CCirc(-30,30){8}{Black}{Purple}
\CCirc(0,75){8}{Black}{Purple}
\Text(0,10)[]{$l_2$}
\Text(-30,65)[]{$l_3$}
\Text(30,65)[]{$l_1$}
\end{picture}
\end{center}
\vspace{1.2cm}
$$
\eqalign{
C_3=
 \sum_{h_i \in {\cal S}'} \int d^4l_1  \delta(l_1^2)
 \delta(l_2^2) \delta(l_3^2)
& M\big((l_1)^{h_1}, i_m, \cdots i_j, (-l_2)^{-h_2} \big)
\cr
&\hspace{-4cm}  \times
M\big((l_2)^{h_2}, i_{j+1} ,\cdots i_l, (-l_3)^{-h_3} \big)
\times
M\big((l_3)^{h_3}, i_{l+1} ,\cdots i_{m-1},  (-l_1)^{-h_1} \big)\,,
\cr}
\hspace{-1cm}\equn
$$ where the summation is over all possible intermediate states.  As
the momentum invariants, $K_1=k_{i_m}+k_{i_m+1}+\cdots k_{i_j}$ etc, are all
non-null, there exist kinematic regimes is which the integration has
non-vanishing support for real loop momentum. In such cases the
remaining one dimensional integral can readily be evaluated
numerically.  In the generic expression of an amplitude the only
integral functions contributing to the triple cut are box functions
and the specific three mass triangle for the cut, 
$$ C_3 = \sum_i c_i
(I_4^i)_{\rm triple-cut} + d_{3m}( I_3^{3m} )_{\rm triple-cut}\,.
\equn
$$ 
The box functions which can contribute are the two-mass-hard,
three-mass and four mass. This equation can be inverted to express
the coefficient $d_{3m}$ in terms of $C_3$ and the box-coefficients.

For the six-point case the cut,
$$\eqalign{
C_3= \sum_{h_i \in {\cal S}'} \int d^4 l_i
& \delta(l_1^2)
\delta(l_2^2) 
\delta(l_3^2)
M_4\big((l_1)^{h_1}, 1, 2 , (-l_2)^{-h_2} \big)
\cr &
\times
M_4\big((l_2)^{h_2}, 3, 4 , (-l_3)^{-h_3} \big)
\times M_4\big((l_3)^{h_3}, 5, 6 , (-l_1)^{-h_1} \big)
\,,}
\equn
$$
only receives contributions from two-mass-hard boxes,
such as $I_4^{2m\; h}[{2,\{3,4\},\{5,6\},1}]$, and the three mass triangle.
The triple cut of a two-mass hard box is,
$$
\eqalign{
(I_4^{2m\; h})_{triple-cut} &=
\int { d^4p
\over p^2 (p-k_2)^2(p-k_2-K_3)^2 (p+k_1)^2 } \bigl|_{\rm cut}
\cr
&=
\int { d^4p  \delta( (p-k_2)^2)  \delta( (p-k_2-K_3)^2) \delta( (p+k_1)^2)
\over p^2  }
\cr
&={ \pi \over 2 (k_1+k_2)^2(k_2+K_3)^2  }\,,
\cr}
\equn
$$
while the triple cut of the three mass triangle is,
$$
\eqalign{
(I_3^{3m})_{triple-cut} &=
\int { d^4p
\over p^2 (p-K_1)^2 (p+K_3)^2 }\bigl|_{\rm cut}
\cr
&=
\int { d^4p  \delta( p^2)  \delta( (p+K_3)^2) \delta( (p-K_1)^2) }
={ \pi \over 2 \sqrt{\Delta_3} }
\cr
&={ \pi \over 2
\sqrt{(K_1^2)^2+(K_2^2)^2+(K_3^2)^2-2( K_1^2K_2^2+K_1^2K_3^2+K_2^2K_3^2)}
}\,.
\cr}
\equn
$$
Thus we see that,
$$
{ \pi \over 2\sqrt{\Delta_3} } d_{3}^{3m}
= C_3 - {\pi \over 2 }\sum_i {  c_{2m\; h,i} \over (k_1+k_2)^2(k_2+K_3)^2 }\,.
\equn
$$

The integral in $C_3$ is well behaved and can be determined
numerically from the tree amplitudes. Using the box-coefficients for
the six-point amplitude~\cite{BDIgravity} we have verified
numerically that,
$$
\eqalign{
 d_{3}^{3m}[ \{ 1^-, 2^- \} , \{ 3^- , 4^+ \} , \{ 5^+, 6^+ \} ] &= 0\,,
\cr
 d_{3}^{3m}[ \{ 1^-, 4^+ \} , \{ 2^- , 5^+ \} , \{ 3^-, 6^+ \} ] &= 0 \, .
\cr}
\equn
$$
The first zero is true for any (massless) gravity theory whilst the
second is true only for $\NeqEight$ supergravity.

For the seven-point amplitude we must also include three-mass boxes
in the triple cut. Using the seven-point box coefficients given in
the appendix we have verified that three mass triangles are absent
in the seven-point {\it NMHV} amplitude. Explicitly,
$$
\eqalign{
d_{3}^{3m}[ \{ 1^-, 2^- \} , \{ 3^- , 4^+ \} , \{ 5^+, 6^+ , 7^+ \} ] &= 0 \,,
\cr
d_{3}^{3m}[ \{ 1^-, 2^- \} , \{ 3^- , 4^+,  5^+  \} , \{ 6^+, 7^+ \} ] &= 0\,,
\cr
d_{3}^{3m}[ \{ 1^-, 2^-, 4^+ \} , \{ 3^- , 5^+   \} , \{ 6^+, 7^+ \} ] &= 0\,,
\cr
d_{3}^{3m}[ \{ 1^-, 4^+ \} , \{ 2^- , 5^+ \} , \{ 3^-, 6^+ , 7^+ \} ] &= 0 \,,
\cr}
\equn
$$
with the first three coefficients vanishing for any matter content
but the last only zero for $\NeqEight$ supergravity.

\subsection{Factorisation}

The unitarity constraints of the previous sections are sufficient to
show the absence of integral functions involving logarithms. This is
sufficient to prove the no-triangle hypothesis for six or fewer
gravitons. At seven-point and beyond, the amplitude may, in
principle, contain rational terms which do not appear in the
four-dimensional cuts.  Unitarity can be used to obtain
these~\cite{vanNeerven:1985xr,BernMorgan,AllEps,Brandhuber2005QCD}
but one must evaluate the cuts fully in $4-2\eps$ dimensions.
Recently, there has been much progress in determining the rational
parts of QCD one-loop amplitudes based on the physical
factorisations of the
amplitudes~\cite{BDKrecursionA,LoopRecursionB,Forde:2005hh}. Gravity
amplitudes are also heavily constrained by factorisation so the
absence of terms other than boxes for six or fewer legs makes it
difficult to envisage their presence at higher points.

More explicitly, consider the multi-particle factorisations.  From
general field theory considerations, amplitudes must factorise (up
to subtleties having to do with infrared singularities) on
multi-particle poles.  For $K^\mu \equiv k_i^\mu+\ldots
+k_{i+r+1}^\mu$ the amplitude factorises when $K$ becomes on shell.
Specifically, as $K^2 \rightarrow 0$ the factorisation properties of
one-loop massless amplitudes are described
by~\cite{BernChalmers},
$$\hspace{-4.9cm}
\eqalign{
M_{n}^{\oneloop}\
& \hskip -.12 cm
 \mathop{\longrightarrow}^{K^2 \rightarrow 0}
\hskip .15 cm
\sum_{\lambda=\pm}  \Biggl[
   M_{r+1}^{\oneloop}\big(k_i, \ldots, k_{i+r-1}, K^\lambda\big) \, {i \over K^2} \,
   M_{n-r+1}^{\tree}\big((-K)^{-\lambda}, k_{i+r}, \ldots, k_{i-1}\big) \cr
& \hskip-1.5cm \null
\!+\!M_{r+1}^{\tree}\big(k_i, \ldots, k_{i+r-1}, K^\lambda\big) {i\over K^2} 
   M_{n-r+1}^{\oneloop}\big((-K)^{-\lambda}, k_{i+r}, \ldots, k_{i-1}\big)
\label{LoopFact} \cr
& \hskip-1.5cm \null
\!+\!M_{r+1}^{\tree}\big(k_i, \ldots, k_{i+r-1}, K^\lambda\big)  {i\over K^2} 
   M_{n-r+1}^{\tree}\big((-K)^{-\lambda}, k_{i+r}, \ldots, k_{i-1}\big) \,
      \cg\,  \Fact_n\big(K^2;k_1, \ldots, k_n\big) \Biggr] ,
\cr}
\hspace{-5cm}\equn
$$
where the one-loop ``factorisation function'' $\Fact_n$ is
helicity independent.

Gravity one-loop amplitudes also have soft and collinear
factorisations. In ref.~\cite{multileg} it was shown that these have
a universal collinear behaviour given by,
$$\hspace{-0.1cm}
M_n^{\rm 1-loop}(\ldots,a^{\lambda_a},b^{\lambda_b},\ldots)
   \mathop{\longrightarrow}^{a \parallel b}\,
\sum_{\lambda}
\!{\rm Split^{\rm gravity}}(z,a^{\lambda_a},b^{\lambda_b})\! \times\!
         M_{n-1}^{\rm 1-loop}(\ldots,P^\lambda,\ldots) \,,
\hspace{-1cm}\equn
$$
when $k_a$ and $k_b$ are collinear. The pure graviton splitting
amplitudes are,
$$
\eqalign{
{\rm Split^{\rm gravity}}_+(z,a^{+},b^{+})\ &=\ 0\,,\cr
{\rm Split^{\rm gravity}}_-(z,a^{+},b^{+})\ &=\ -{ 1 \over z(1-z) }
                 { \spb{a}.{b} \over \spa{a}.b }\, ,\cr
 {\rm Split^{\rm gravity}}_+
(z,a^{-},b^{+})\ &=\ - { z^3 \over 1-z }
                 { \spb{a}.{b} \over \spa{a}.b } \,. \cr}
\equn
$$
There is also a universal soft behaviour given by,
$$
M_n^{\rm 1-loop}(\ldots,a,s^\pm,b,\ldots)\ \mathop{\longrightarrow}^{k_s\to0}\
 {\cal S}^{\rm gravity} (s^\pm)
\times
   M_{n-1}^{\rm 1-loop}(\ldots,a,b,\ldots)\,,
\equn
$$
when $k_s$ becomes soft. For the limit $k_n \rightarrow 0$ in
$M_n^\tree(1,2,\ldots,n)$, the gravitational soft factor (for
positive helicity) is,
$$
\hspace{1.5cm}\eqalign{
\Soft_n\ \equiv\
\SoftGrav(n^+)
\ &=\ { -1 \over \spa{1}.{n} \spa{n,}.{n-1} }
  \sum_{i=2}^{n-2} { \spa{1}.{i} \spa{i,}.{n-1} \spb{i}.{n}
                                         \over \spa{i}.{n} }\, .\cr}
\equn\label{GravTreeSoftFactor}
$$

Note that the collinear behaviour is only a ``phase singularity''
for real momenta~\cite{multileg}, however it should be regarded as a
genuine singularity when using complex momenta.

These factorisations place constraints on the rational terms $R_n$.
Since $R_n=0$ for $n \leq 6$ the natural solution is $R_n=0$ for all
$n$.  For QCD the factorisation constraints have been turned into
recursion relations for the rational
terms~\cite{BDKrecursionA,LoopRecursionB}. If this bootstrap also
applies to gravity amplitudes then we would be able to immediately
deduce that $R_n=0$ for $\NeqEight$ amplitudes. At present a direct
calculation of the rational terms beyond six-points seems unfeasible
although there has been recent progress in producing algorithms
focused on computing the rational terms in six-point QCD
amplitudes~\cite{Xiao:2006vt,Binoth:2006hk}.

\section{Checking Bubble-cuts by Large-$z$ Shifts}

In this section we look at a different way to test for
bubble integral functions in the two-particle cuts. This approach is
based on the scaling behaviour of amplitudes under specific shifts 
of the loop momenta. 

Starting from equations~(\ref{cut}) and~(\ref{discontinuities}),
lifting the integral implies,
$$
\eqalign{
&\hspace{-4cm}
\Mtree\big( (-l_1)^{-},i,\cdots, j,(l_2)^{-}\big)\times\Mtree\big( (-l_2)^{+},j+1,\cdots, i-1,(l_1)^{+}\big)\cr
 &\hspace{-3cm}=\sum_{i\in \cal C'} { c_{i} \over (l_1-K_{i,4})^2 (l_2-K_{i,2})^2 }
+\sum_{j\in\,\cal D'} {d_{j} \over (l_1-K_{j,3})^2 }
+ e_{k'}+ D(l_1,l_2)\,.
}
\hspace{-4cm}\equn\label{heurcut}
$$
Here $D(l_1,l_2)$ is a total derivative, $\int \dlips(-l_1,l_2)
D(l_1,l_2) =0$, which may or may not be present.
Note that in the above a number of boxes and
triangles may contribute but only one bubble. Let us consider
 (\ref{heurcut}) under the shift of the two-cut legs,
$$
\lambda_{l_1} \longrightarrow \lambda_{l_1} +z\, \lambda_{l_2}\,,
\;\;\;\;
\tilde\lambda_{l_2} \longrightarrow \tilde\lambda_{l_2} 
-z\, \tilde\lambda_{l_1}\,.
\equn\label{zshifteq}
$$
This shift does not change the coefficients but it does enter the propagator
terms (and possibly the $D(l_1,l_2)$). In the large-$z$ limit the
propagators will vanish as $1/z$ leaving behind the
bubble coefficient $e_{k'}$. This suggests a test for bubble terms: if,
$$
\lim_{z\rightarrow\infty} \Mtree\big( (-l_1)^{-h_1},i,\cdots,
j,(l_2)^{h_2}\big) \times\Mtree\big( (-l_2)^{-h_2},j+1,\cdots, i-1,(l_1)^{h_1}\big)
{\longrightarrow 0}\,, \equn
$$
in the large-$z$ limit, then,
$$
e_{k'} = 0\,,
\equn
$$ under the assumption that the total derivative vanishes at
infinity.  In the following section we discuss criteria for
when this test may be used. 
This test is particularly useful as in many cases it follows from the 
general behaviour of gravity tree amplitudes and may be tested numerically 
when the tree amplitudes are known. 

\subsection{Relation to large $t$ behaviour} 

A key step is to relate the large $z$ behaviour to the large
$t$ behaviour of the cut parameterised as in the previous section.
In that section, following~\cite{BrittoN=1,Britto:2006sj}, we
discussed how the integral functions that a given term in a unitarity cut contributes to are
determined by the power, $n$, of $1/(P_{a\dot
a}\lambda^a\tilde\lambda^{\dot a})^n$. A term in the cut integral
(\ref{cutintB}) can be written as a rational
expression in holomorphic and anti-holomorphic spinors $\lambda^a$
and $\tilde\lambda^{\dot a}$ respectively (recall these spinors are
NOT the same as the $\lambda_{l_i}$ but are related via
$l_1=t\lambda\tilde\lambda$),
$$
\int
\frac{\spa{\lambda}.{d\lambda}\spb{\tilde\lambda}.{d\tilde\lambda}}{(P_{a\dot
a}\lambda^a\tilde\lambda^{\dot a})^{n}}\, \frac{n_{a_1\ldots
a_j,\dot a_1\ldots \dot
a_k}\lambda^{a_1}\cdots\lambda^{a_j}\tilde\lambda^{\dot
a_1}\cdots\tilde\lambda^{\dot a_k} }{d_{b_1\ldots b_l,\dot b_1\ldots
\dot b_m}\lambda^{b_1}\cdots\lambda^{b_l}\tilde\lambda^{\dot
b_1}\cdots\tilde\lambda^{\dot b_m}}, \equn
$$
where the tensors $n_{a_1\ldots a_j,\dot a_1\ldots \dot a_k}$ and
$d_{b_1\ldots b_l,\dot b_1\ldots \dot b_m}$ contain no factors of
$(P_{a\dot a}\lambda^a\tilde\lambda^{\dot a})$.

Since the integrand must carry spinor weight $-2$ in $\lambda$ and
$\tilde\lambda$, the counters $j,k,l,m$ and $n$ obey $j-l-n=-2$
and $k-m-n=-2$. The $n_{a_1...\dot a_k}$ are non-vanishing for the
contractions,
$$
n_{a_1\ldots a_j}\,^{\dot a_1\ldots \dot a_k}\lambda^{a_1}\cdots\lambda^{a_j}
\big(\lambda^{b_1}P_{b_1\dot a_1}\big)\cdots \big(\lambda^{b_k}P_{b_k\dot a_k}\big)\neq 0\, ,
\equn
$$
and similarly for $d_{b_1...\dot b_m}$. This can always be achieved: were
the above contraction to vanish for all values of
$\lambda$, then the spinor obtained by contracting all but one index
has to be parallel to $\lambda^{b_k}P_{b_k}\,^{\dot a_k}$, that is,
$$
n_{a_1\ldots a_j}\,^{\dot a_1\ldots \dot
a_k}\lambda^{a_1}\cdots\lambda^{a_j} \big(\lambda^{b_1}P_{b_1\dot
a_1})\cdots \big(\lambda^{b_{k-1}}P_{b_{k-1}\dot a_{k-1}}\big)
=n'(\lambda,P\lambda)\big(\lambda^{b_k}P_{b_k}\,^{\dot a_k}\big),
\equn
$$
with $n'$ a tensor of lower rank in $\lambda$ and $\tilde\lambda$.
We would then be able to pull out a power of $(P_{a\dot
a}\lambda^a\tilde\lambda^{\dot a})$ and write the contraction of $n$ as,
$$
n(\lambda,\tilde\lambda) = n'(\lambda,\tilde\lambda)\,(P_{a\dot
a}\lambda^a\tilde\lambda^{\dot a}),
\equn
$$
contrary to our condition that no such factors exist.

The central observation is that the power $n$ of $1/(P_{a\dot
a}\lambda^a\tilde\lambda^{\dot a})^n$ is related to the leading
power in large-$z$ of $\lambda_{l_i}$. The shift in~(\ref{zshifteq}) translates into a shift on the
$\lambda$ and $\tilde\lambda$ of,
$$
\lambda_a \longrightarrow \lambda_a\,,\quad \tilde\lambda_{\dot
a}\rightarrow \tilde\lambda_{\dot a}+z \,\lambda^{a}P_{a\dot a}\,.
\equn\label{onshellshift}
$$
The terms $(P_{a\dot a}\lambda^a\tilde\lambda^{\dot a})$ are
invariant under the shift, so the leading term at large-$z$ is
given by,
$$
z^{k-m}\frac{1}{(P_{a\dot a}\lambda^a\tilde\lambda^{\dot a})^{n}}\times
\frac{n(\lambda,P\lambda)}{d(\lambda,P\lambda)}.
\equn
$$
Using $n=(k-m)+2$ one finds that the large-$z$ scaling is,
$$
\sim z^{n-2},
\equn
$$ 
for a term with a $1/(P_{a\dot a}\lambda^a\tilde\lambda^{\dot
a})^n$ factor.  Consequently, if the product of the two tree
amplitudes vanishes as $z\longrightarrow\infty$ then this product can
only be composed of terms with $n\leq 1$. These terms do not
contribute to bubble functions and hence the coefficient of the bubble
corresponding to this cut must vanish.~\footnote{Note that since there is only
a single bubble in each cut,  there is no possibility of
cancellation. It is not uncommon for cancellations to occur
amongst the box functions appearing in a cut. The propagators of a
single box vanish as $1/z^2$, however, as in many of the cases we
consider, the leading terms cancel amongst
the boxes leaving a $1/z^4$ behaviour as $z\longrightarrow\infty$.}

\subsection{Using the large-$z$ test for Gravity Amplitudes.}
\label{usingthelargeztest}

In this section we apply the test of the previous section to
the two-particle cuts for graviton scattering in $\NeqEight$
supergravity. We can use the behaviour of the gravity amplitudes
under the shift~(\ref{zshifteq}) to determine the behaviour of the
cut. We will need to consider two types of cut: singlet cuts where
only graviton amplitudes are needed and non-singlet cuts where
amplitudes involving other  states in the supergravity multiplet contribute.

It is useful to briefly review the known results for the large-$z$
behaviour of gravity amplitudes under the shifts,
$$
\lambda_i\rightarrow\lambda_i+z\,\lambda_j\,,
\quad\tilde\lambda_j\rightarrow\tilde\lambda_j-z\tilde\lambda_i.\equn\label{lzshift}
$$
The scaling of a given tree amplitude depends on the helicity of the
two shifted legs and the helicity of the scattering gravitons. For
{\it MHV}-amplitudes there is an explicit all-$n$
representation of the tree amplitudes~\cite{BerGiKu}. This can be used to show 
that~\cite{BBSTgravity,CSgravity,gravCSW},
$$
\eqalign{
(h_i,h_j)=&  (+,+), (-,-),(+,-) :  
\Mtree|_{z\longrightarrow \infty}
\sim {1\over z^2 }, \cr
(h_i,h_j)=&  (-,+)\quad\quad\quad\quad\quad\quad \;\; :  
\Mtree|_{z\longrightarrow \infty}\sim { z^6 }.
}\equn\label{scaleequ}
$$
Slightly more surprisingly this behaviour extends to {\it NMHV}
amplitudes also - at least up to seven points where we have checked
the result explicitly. It is tempting to conjecture that this is
true for all graviton tree amplitudes. We only need the behaviour up
to seven points to test for bubbles in the six and seven point
amplitudes.

We first
consider the {\it MHV} case for arbitrary $n$.
The singlet cuts are of the form,
$$
\Mtree\big((l_1)^+,1^-,2^-,3^+,\cdots r^+, (l_2)^+)\times
\Mtree\big((l_1)^-,(r+1)^+,\cdots n^+, (l_2)^-). \equn
$$
When we shift this we find that each tree shifts as $1/z^2$, so the product
behaves as $1/z^4$ at large-$z$ and we can deduce that the bubble
integral function $I_2( K_{1\ldots r})$ has vanishing coefficient.
The non-singlet cut is more involved,
$$\hspace{-0.2cm}
\eqalign{ & \sum_h  \Mtree\big((-l_1)^{-h},2^-,3^+,\cdots r^+, (l_2)^{h}
)\!\times\! \Mtree\big((-l_2)^{-h},(r+1)^+,\cdots n^+, 1^-, (l_1)^{h}\big) =\cr &
\Mtree\big((-l_1)^{-},2^-,3^+,\cdots r^+, (l_2)^{+} \big)\times
\Mtree\big((-l_2)^{-},(r+1)^+,\cdots n^+,1^-, (l_1)^{+}\big)\cr & \hspace{9.3cm}\times 
\sum_{h\in{\cal S}'} \left( {
\spa{2}.{l_1} \spa{1}.{l_2}     \over \spa{2}.{l_2} \spa{1}.{l_1}}
\right)^{2h-4} \cr & = \rho \times \Mtree\big((-l_1)^{-},2^-,3^+,\cdots r^+,
(l_2)^{+} \big)\times \Mtree\big((-l_2)^{-},(r+1)^+,\cdots n^+,1^-, (l_1)^{+}\big),\cr}
\hspace{-4cm}\equn\label{eqgodknows}
$$
where,
$$
\rho =  \left({
\spa{2}.{l_2} \spa{1}.{l_1} - \spa{2}.{l_1} \spa{1}.{l_2} 
\over
\spa{2}.{l_1} \spa{1}.{l_2}  }\right)^8
=\left({
\spa{l_1}.{l_2} \spa{1}.{2}  
\over
\spa{2}.{l_1}  \spa{1}.{l_2} }\right)^8.
\equn
$$
Under the shift the amplitudes scale as,
$$
\eqalign{&\Mtree\big((-l_1)^{-},2^-,3^+,\cdots r^+, (l_2)^{+} \big)
\,\sim\,   z^6\,,\cr
&\Mtree\big((-l_2)^{-},(r+1)^+,\cdots n^+,1^-, (l_1)^{+}\big)\,\sim\,   1/z^2\,,
}
\equn
$$
however the $\rho$-factor scales, noting that $\spa{l_1}.{l_2}$ is unshifted, as,
$$
\rho\,\sim\,   {1 \over z^8},
\equn
$$
and we find the non-singlet cuts scale as $1/z^4$, exactly as in the
singlet case.  Within the sum over the multiplet~(\ref{eqgodknows})
the product of tree amplitudes scales as $z^4$ for any given 
state and the simplification only arises when the sum over the entire
$\NeqEight$ multiplet is performed.

We will now discuss the possible cuts of
the six and seven-point {\it NMHV} amplitudes.
For any singlet cut, 
$$
\eqalign{
\Mtree\big((-l_1)^-,\ldots ,(l_2)^- \big)
\times
\Mtree\big((-l_2)^+,\ldots ,(l_1)^+\big)\,,}
\equn
$$
the trees both vanish as $1/z^2$
under the shift~(\ref{lzshift})
and we deduce that bubble integral functions are
absent from these cuts. 
Thus bubbles corresponding to singlet cuts are absent 
up to seven-points. 

The non-singlet cuts are more involved. For the six-point amplitude,
\linebreak  $M(1^-,2^-,3^-,4^+,5^+,6^+)$
there are two types of cut corresponding to the cuts $C_{234}$ and $C_{34}$.
For the $C_{234}$ cut the amplitudes are a product of an {\it MHV} and
a $\overline{\it MHV}$. Summing over the multiplet gives an overall $\rho$ factor just as in the ${\it MHV}$ case and we deduce the coefficient of this bubble function is absent. The $C_{34}$ cut
is given by, 
$$
\sum_{h\in\,{\cal S}'}
\Mtree\big((-l_1)^{-h},3^-,4^+,(l_2)^h \big)
\times
\Mtree\big((-l_2)^{-h},1^-,2^-,5^+,6^+,(l_1)^h\big)\,.
\equn
$$
The amplitude involving a state of helicity $h$ behaves as,
$$
\Mtree\big((-l_1)^{-h},1^-,2^-,5^+,6^+,(l_2)^h\big)
\,\sim\, z^{2h+2}\,,
\equn
$$
which is a natural refinement of~(\ref{scaleequ}) and can be checked using the
form of the amplitude in appendix~\ref{sixpointmultiplet}. 
Thus
the product of the two tree amplitudes in the cut (which will have
states of $\pm h$) will always behave as $z^4$ and the corresponding 
scattering amplitude will contain bubble functions (or boundary
terms). By explicit computation it can be seen that after 
including all the states from the $\NeqEight$ supergravity multiplet 
we have,
$$\hspace{-0.7cm}
\sum_{\rm \ \ \ \ \ {\cal N}=8\, multiplet}\!\!\!\!\!\!\!\!\!\!\!\!\!\!
\Mtree\big((-l_1)^{-h},3^-,4^+,(l_2)^h \big)
\times
\Mtree\big((-l_2)^{-h},1^-,2^-,5^+,6^+,(l_1)^h\big)|_{z\longrightarrow\infty}\,
\longrightarrow 0  \,,
\equn
$$ 
and the bubble functions drop out.  This calculation shows how
the sum over the multiplet leads to the absence of bubble
functions in the six-point {\it NMHV} amplitude even though they 
are present in the contribution from any single state in the multiplet. 
For the seven-point amplitude 
$M(1^-,2^-,3^-,4^+,5^+,6^+,7^+)$
there are three types of cut: $C_{234}$, $C_{345}$ 
and $C_{34}$. Of these, the large $z$ behaviour of 
$C_{234}$ and $C_{345}$ can be checked using the 
six-point amplitudes verifying the absence of these bubbles. 

\section{Consequences and Conclusions}
In this paper we have given further evidence that the one-loop
perturbative expansion for $\NeqEight$ supergravity is much closer
to that of $\NeqFour$ super Yang Mills than expected from power
counting arguments. We argue that the one-loop amplitudes are
composed entirely of box integral functions and contain
``no-triangle'' (or bubble or rational) integral functions. We have
provided evidence rather than a proof for this ``no-triangle
hypothesis'', but the evidence amounts to a proof for the six-point
amplitudes.  The evidence for $n$-point amplitudes with $n\geq7$
based on unitarity, factorisation and IR behaviour is, for us,
compelling.

The cancellation we observe  is {\it not} ``diagram-by-diagram'' -
at least not in any computational framework we are aware of.
Individual diagrams appear to have loop momentum polynomials of
degree $2(r-4)$ and simplification only occurs when the diagrams are
summed. The simplification observed is quite dramatic: to yield only
boxes the simplification would be equivalent to a cancellation
between terms such that the leading $(r-4)$ terms in the loop
momentum polynomials cancel.

The ``no-triangle hypothesis'' applies strictly to one-loop
amplitudes only. However we expect it to have consequences for
higher loops. For $\NeqEight$ supergravity in $D=4$ the four point
amplitude is expected to diverge at five loops~\cite{BDDPR}. This
argument is based on power counting and the known symmetries of the
theory~\cite{HoweStelle}. Specifically, the argument attempts to
estimate the power of the loop momentum integral of individual
higher loop diagrams and finds that they generically have twice the
power of the equivalent Yang-Mills diagram. Cancellations between
diagrams analogous to those occuring at  one-loop would lead to a softer
UV behaviour than this prediction with the theory possibly even
being finite.

Presumably there is a symmetry underlying this simplification.  We
are not aware of any potential candidates for this symmetry. 
Although examining on-shell
amplitudes has many advantages, the nature of the underlying symmetry
is obscure in the amplitudes. The symmetries implied by the twistor
duality~\cite{WittenTopologicalString} are one potential source,
although originally the duality seemed to involve super-conformal
rather than Einstein gravity~\cite{BerkovitsWitten}. Recently
twistor strings involving Einstein gravity have been
constructed~\cite{Abou-Zeid:2006wu} and it would be interesting to
explore these for potential symmetries. If $\NeqEight$ were
``weak-weak'' dual to a UV finite string theory then obviously the
finiteness of $\NeqEight$ supergravity would follow.


\vspace{1.0cm} \noindent{\bf Acknowledgments}

We thank Zvi Bern for many useful discussions. This research was
supported in part by the PPARC and the EPSRC of the UK and in part
by grant DE-FG02-90ER40542 of the US Department of Energy.

\vfill\eject

\appendix

\section{Seven-Point Amplitude}
\label{SevenPointAppendix}

The seven-point {\it NMHV} amplitude $M_7(1^-,2^-,3^-,4^+,5^+,6^+,7^+)$ can be
expressed as a sum of scalar boxes together with rational
coefficients;
$$
\Mloop= \sum_{a}  c_a I_4^a\,.\equn\label{OnlyBoxesEQ}
$$
The scalar boxes can be of four types, three mass, two-mass-hard,
two-mass-easy and one mass shown below with our choice of labelling.

%

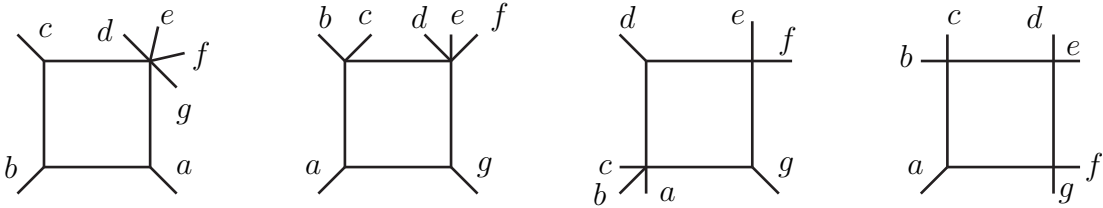
\begin{center}
\begin{picture}(110,80)(0,0)
\SetWidth{1}
\Line(10,10)(10,50)
\Line(10,50)(50,50)
\Line(50,50)(50,10)
\Line(50,10)(10,10)
\Line(10,10)(0,0)
\Line(10,50)(0,60)
\Line(50,50)(40,60)
\Line(50,50)(53,63)
\Line(50,50)(63,53)
\Line(50,50)(60,40)
\Line(50,10)(60,0)
\Text(60,10)[l]{$ a$}
\Text(-5,10)[l]{$ b$}
\Text(8,62)[l]{$ c$}
\Text(30,62)[l]{$ d$}
\Text(54,67)[l]{$ e$}
\Text(66,52)[l]{$ f$}
\Text(60,30)[l]{$ g$}
\end{picture}
\begin{picture}(110,80)(0,0)
\SetWidth{1}
\Line(10,10)(10,50)
\Line(10,50)(50,50)
\Line(50,50)(50,10)
\Line(50,10)(10,10)
\Line(10,10)(0,0)
\Line(10,50)(0,60)
\Line(10,50)(20,60)
\Line(50,50)(60,60)
\Line(50,50)(40,60)
\Line(50,50)(50,60)
\Line(50,10)(60,0)
\Text(60,10)[l]{$ g$}
\Text(-5,10)[l]{$ a$}
\Text(0,67)[l]{$ b$}
\Text(15,67)[l]{$ c$}
\Text(35,67)[l]{$ d$}
\Text(50,67)[l]{$ e$}
\Text(65,67)[l]{$ f$}
\end{picture}
\begin{picture}(110,80)(0,0)
\SetWidth{1}
\Line(10,10)(10,50)
\Line(10,50)(50,50)
\Line(50,50)(50,10)
\Line(50,10)(10,10)
\Line(10,10)(0,0)
\Line(10,10)(10,0)
\Line(10,10)(0,10)
\Line(10,50)(0,60)
\Line(50,50)(50,65)
\Line(50,50)(65,50)
\Line(50,10)(60,0)
\Text(60,10)[l]{$ g$}
\Text(15,0)[l]{$ a$}
\Text(-10,0)[l]{$ b$}
\Text(-8,10)[l]{$ c$}
\Text(0,67)[l]{$ d$}
\Text(42,67)[l]{$ e$}
\Text(60,58)[l]{$ f$}
\end{picture}
\begin{picture}(80,80)(0,0)
\SetWidth{1}
\Line(10,10)(10,50)
\Line(10,50)(50,50)
\Line(50,50)(50,10)
\Line(50,10)(10,10)
\Line(10,10)(0,0)
\Line(10,50)(0,50)
\Line(10,50)(10,60)
\Line(50,50)(50,60)
\Line(50,50)(60,50)
\Line(50,10)(50,0)
\Line(50,10)(60,10)
\Text(52,0)[l]{$ g$}
\Text(-5,10)[l]{$ a$}
\Text(-8,52)[l]{$ b$}
\Text(10,67)[l]{$ c$}
\Text(40,67)[l]{$ d$}
\Text(55,55)[l]{$ e$}
\Text(62,10)[l]{$ f$}
\end{picture}
\end{center}

The coefficients of the box-functions can be obtained by
unitarity~\cite{BDDKa,BDDKb}. Recently, it was observed that the
box-coefficients can be efficiently obtained from the quadruple
cut~\cite{BrittoUnitarity},

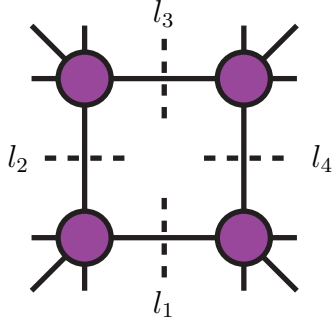
\begin{center}
\begin{picture}(100,100)(0,0)
\SetWidth{2}
\Line(20,20)(80,20)
\Line(80,20)(80,80)
\Line(80,80)(20,80)
\Line(20,80)(20,20)
\Line(20,20)(0,0)
\Line(20,20)(0,20)
\Line(20,20)(20,0)
\Line(80,20)(100,0)
\Line(80,20)(80,0)
\Line(80,20)(100,20)
\Line(80,80)(100,100)
\Line(80,80)(100,80)
\Line(80,80)(80,100)
\Line(20,80)(0,100)
\Line(20,80)(20,100)
\Line(20,80)(0,80)
\CCirc(20,20){10}{Black}{Purple}
\CCirc(20,80){10}{Black}{Purple}
\CCirc(80,20){10}{Black}{Purple}
\CCirc(80,80){10}{Black}{Purple}
\DashLine(50,95)(50,65){4}
\DashLine(50,5)(50,35){4}
\DashLine(5,50)(35,50){4}
\DashLine(65,50)(95,50){4}
\Text(50,-5)[]{$l_1$}
\Text(50,105)[]{$l_3$}
\Text(-5,50)[]{$l_2$}
\Text(110,50)[]{$l_4$}
\end{picture}
\end{center}
which yields the coefficient of the corresponding integral function,
$$
\eqalign{
 c={ 1 \over 2 } \sum_{h_i\in\,\cal S}
  \biggl( \Mtree\big( &(-\ell_1)^{-h_1},i_1,  \ldots,i_2,(\ell_2)^{h_2}\big) \times
  \Mtree\big( (-\ell_2)^{-h_2},i_3,\ldots,i_4,(\ell_3)^{h_3}\big)\cr
&\hspace{-2cm}\times \Mtree\big( (-\ell_3)^{-h_3},i_5,\ldots,i_6,(\ell_4)^{h_4}\big) \times
 \Mtree\big( (-\ell_4)^{-h_4},i_7,\ldots,i_8,(\ell_1)^{h_1}\big) \biggr)\,.
\cr}
\hspace{-1.5cm}\equn
$$ 
In this expression the sum is over all possible states of the
$\NeqEight$ multiplet and all possible helicity configurations for
which the four tree amplitudes are non-zero. The four cut momenta
are all on-shell, $l_i^2=0$. If the four tree amplitudes have four or
more legs then this is solved for real momenta whereas if a corner
has only three legs then the solution involves complex momenta.
Alternately the box-coefficients of $\NeqEight$ can be obtained from
those of $\NeqFour$ super Yang-Mills where for example, with the
above labelling, the coefficients of the three-mass boxes are
related by,
$$\hspace{1cm}
\eqalign{
c_{N=8}^{[a,\{b,c\},\{d,e\},\{f,g\}]} 
 = 
2 s_{bc}s_{de}s_{fg}\times  &
c_{N=4}^{[a,\{b,c\},\{d,e\},\{f,g\}]}\times
c_{N=4}^{[a,\{c,b\},\{e,d\},\{g,f\}]}.
\cr}
\equn
$$

To implement the quadruple cuts requires a knowledge of the tree
amplitudes up to and including six-points,  where two
particles are states other than gravitons. The three, four and five
points amplitudes are all {\it MHV} or $\overline{\it MHV}$ amplitudes and relatively simple.
For {\it MHV} amplitudes the amplitude with $n-2$ gravitons and two
non-graviton particles are related to the {\it MHV} amplitude by,
$$\hspace{-0.05cm}
M(1_h^-,2^-,3^+,\cdots,(n-1)^+, n^+_h)
=
\left( { \spa{2}.{n} \over \spa{2}.1 }\right)^{2h-4}
\!\!\!M(1^-,2^-,3^+,\cdots,(n-1)^+, n^+).
\hspace{-1cm}\equn
$$
For the six-point corners the tree may be {\it MHV} or {\it NMHV}. The
six-graviton {\it NMHV} tree amplitudes were computed
recently~\cite{CSgravity,BDIgravity}. To complete the calculation of
the box-coefficients we also need the six-point amplitudes with
two non-gravitons. These are presented in
appendix~\ref{sixpointmultiplet}.

\subsection{Definitions}
 The coefficients of the boxes are expressed using spinor products.
 We use the notation $\spa{j}.{l}\equiv \langle
 j^- | l^+ \rangle $, $\spb{j}.{l} \equiv \langle j^+ |l^- \rangle
 $, with $| i^{\pm}\ra $ being  massless Weyl spinors with momentum
 $k_i$ and chirality $\pm$~\cite{SpinorHelicity}. The
 spinor products are related to momentum invariants by
 $\spa{i}.j\spb{j}.i=2k_i \cdot k_j\equiv s_{ij}$ .
 As in twistor-space studies we use the notation,
 $$
 \lambda_i \;=\; |  i^+\ra   %
 \; , \;\;\;
 \tilde\lambda_i \;= \; |  i^-\ra  %
 \,.
 \equn
 $$
 We also define spinor strings,
 $$
 \eqalign{
 \BBRQ k {{K}_{i\ldots j}} l \;& \equiv\;
 \BRQ {k^+} {\Slash{K}_{i\ldots j}} {l^+} \;\equiv\; \BRQ
 {l^-} {\Slash{K}_{i\ldots j}} {k^-} \;\equiv\;
 \la l |{{K}_{i\ldots j}}| k] \;\equiv\; \sum_{a=i}^j\spb k.a\spa a.l\,,
 \cr
 \la k | K_{i\ldots j} K_{m\ldots n} | l \ra &
 \equiv\; \BRQ {k^-}
 {\Slash{K}_{i\ldots j}\Slash{K}_{m\ldots n}}
 {l^+}= \sum_{a=i}^j\sum_{b=m}^n\spa k.a\spb a.b \spa b.l \,,
 \cr
  {[k |{{K}_{i\ldots j}{K}_{m\ldots n}} |l]} &  \equiv\;
 {\BRQ {k^+} {\Slash{K}_{i\ldots j}\Slash{K}_{m\ldots n}} {l^-}} \;\equiv\;
 \sum_{a=i}^{j}\sum_{b=m}^{n}\spb k.a\spa a.b\spb b.l\,,
 \cr}
 \equn
 $$
etc. We will often use the momentum invariants
$s_{ij}=(k_i+k_j)^2$ and $t_{ijk}=(k_i+k_j+k_k)^2$.

\subsection{Three Mass Boxes}

The three mass boxes have one graviton attached to one corner and
two gravitons to each of the others. The three-point corner is
$\overline{\it MHV}$ while the four-point corners are {\it MHV}. This means
that all four corners are relatively simple and that different
helicity configurations are also relatively simply related. In the
case of $L_6$ there is a summation over the full $\mathcal N=8$
multiplet running in the loop. We get,
$$
\eqalign{
&c{[a^+,\{ b^+,c^+ \} , \{ d^-,e^+ \} , \{f^-,g^-\} ]}
=
L_0,
\cr
& c{[a^+,\{ b^+,c^+ \} , \{ d^-,e^- \} , \{f^-,g^+\} ]}
= L_1 =
\left( {\la f | K_{de}K_{bc} | a \ra
\over
\BRi{e}{a}{bc} \spa{f}.g } \right)^8 L_0,
\cr
& c{[a^+,\{ b^+,c^- \} , \{ d^+,e^+ \} , \{f^-,g^-\} ]}
= L_2=  \left({  \spa{a}.c \spb{d}.e \over \BRi{e}{a}{bc} } \right)^8 L_0
\cr
& c{[a^-,\{ b^+,c^+ \} , \{ d^-,e^- \} , \{f^+,g^+\} ]}
= L_3 =\left({\BRT{a}{bc}{fg}{a}
\over \BRi{e}{a}{bc} \spa{f}.g
} \right)^8  L_0,
\cr
& c{[a^-,\{ b^-,c^- \} , \{ d^+,e^+ \} , \{f^+,g^+\} ]}
= 0,
\cr
& c{[a^-,\{ b^+,c^- \} , \{ d^+,e^+ \} , \{f^+,g^- \} ]}
= L_4 =\left({\
\spa{a}.c \spa{a}.g \spb{d}.e \over \BRi{e}{a}{bc} \spa{f}.g
} \right)^8  L_0,
\cr
& c{[a^-,\{ b^-,c^+ \} , \{ d^-,e^+ \} , \{f^+,g^+ \} ]}
= L_5 =\left({ \BRi{e}{a}{fg} \spa{a}.b
\over \BRi{e}{a}{bc} \spa{f}.g
} \right)^8   L_0,
\cr
& c{[a^+,\{ b^-,c^+ \} , \{ d^-,e^+ \} , \{f^-,g^+ \} ]}
= L_6
\cr
& \hskip 2.0 truecm
=
\left(
\spa{b}.a   [e|K_{fg}|a\ra \la f | K_{de}K_{bc},a]-
\spa{f}.a  [e|K_{bc}|a\ra \la b | K_{de}K_{fg},a]
\over
\BRi{e}{a}{bc}
\spa{f}.{g}
\la  a | K_{bc} K_{de} | a \ra
  \right)^8
L_0,
\cr}
\hspace{-1.5cm}\equn
$$
where,
$$
\eqalign{
L_0 = &
{- s_{bc}s_{de}s_{fg}   \spa{g}.f^6 \BRi{e}{a}{bc}^8  ( t_{abc}t_{fga}-s_{bc}s_{fg} )^2
\over
2 \spb{d}.e^2 \spa{b}.c^2{ \prod \atop {x=b,c,g,f} }\spa{a}.x
{ \prod \atop {y=d,e} } \BRi{y}{a}{fg} \BRi{y}{a}{bc}
{ \prod \atop {z=b,c} }\BRT{z}{de}{fg}{a}
{ \prod \atop {w=f,g} } \BRT{w}{de}{bc}{a}
}.
\cr}
\equn$$
\newpage

\subsection{Two Mass Hard Boxes}

The two mass hard boxes have two adjacent three-point corners, a
four-point corner and a five-point corner. The four- and five-point
corners are {\it MHV} and of the two three-point corners one is {\it MHV}
and the other is $\overline{\it MHV}$. The two ways of assigning these
give rise to the $G_i$ and $H_i$ terms below. Because all corners
are either {\it MHV} or $\overline{\it MHV}$, the different helicity
configurations are simply related. We get,

$$\hspace{-0.5cm}
\eqalign{
c[a^-,\{b^-,c^-\}, \{d^+,e^+,f^+\},g^+ ]
&= G_0,
\cr
c[a^-,\{b^-,c^+\}, \{d^-,e^+,f^+\},g^+ ]
&=
G_1+H_1=
\left({ [ c | K_{abc} | d \ra \over t_{abc} } \right)^8
G_0+\left({  \spa{a}.b [ g | K_{abc} | d \ra \over \spa{b}.c t_{def} }  \right)^8 H_0,
\cr
c[a^-,\{b^+,c^+\}, \{d^-,e^-,f^+\},g^+ ]
&=G_2+H_2=
\left({ \spb{b}.c \spa{d}.e \over t_{abc} } \right)^8
G_0+\left({  \spa{d}.e [ g | K_{abc} | d \ra \over \spa{b}.c
 t_{def} }  \right)^8 H_0,
\cr
c[a^+,\{b^+,c^+\}, \{d^-,e^-,f^-\},g^+ ]
&= 0,
\cr
c[a^+,\{b^-,c^-\}, \{d^-,e^+,f^+\},g^+ ]
&=G_3+H_3=
\left({ [a | K_{abc} | d \ra \over t_{abc} } \right)^8
G_0+\left({ [ g | K_{abc} | d \ra \over t_{def} } \right)^8 H_0,
\cr
c[a^+,\{b^-,c^+\}, \{d^-,e^-,f^+\},g^+ ]
&=G_4+H_4=
\left({ \spb{a}.c \spa{d}.e  \over t_{abc} } \right)^8
G_0
+\left(
{  \spa{d}.e [ g | K_{abc} | b \ra \over \spa{b}.c t_{def}}
\right)^8 H_0,
\cr
c[a^+,\{b^-,c^-\}, \{d^+,e^+,f^+\},g^- ]
&=G_5+H_5=
\left({ [a | K_{abc} | g \ra \over t_{abc} } \right)^8
G_0+ H_0,
\cr
c[a^+,\{b^-,c^+\}, \{d^-,e^+,f^+\},g^- ]
&=G_6+H_6=
\left(
{  \spa{d}.{g} \spb{a}.b  \over  t_{abc} }
\right)^8
G_0+
\left(
{  \la d | K_{def} K_{abc}  | b \ra
\over   \spa{b}.c  t_{def}  }
\right)^8 H_0,
\cr
c[a^+,\{b^+,c^+\}, \{d^-,e^-,f^+\},g^- ]
&=G_7+H_7=
0.
G_0+\left({ t_{abc} \spa{d}.e \over \spa{b}.c  t_{def} }  \right)^8 H_0,
\cr
c[a^-,\{b^-,c^+\}, \{d^+,e^+,f^+\},g^- ]
&=G_8+H_8=
\left({   [ c | K_{abc} | g \ra \over t_{abc} } \right)^8
G_0+\left({ \spa{a}.b \over \spa{c}.b }   \right)^8 H_0,
\cr
c[a^-,\{b^+,c^+\}, \{d^-,e^+,f^+\},g^- ]
&=G_9+H_9=
\left({ \spa{g}.d \spb{b}.c \over t_{abc} }\right)^8
G_0+\left( {  \la a | K_{bc} K_{def} | d \ra \over \spa{b}.c t_{def} } \right)^8 H_0,
\cr}
\equn$$
where,
$$
\eqalign{
G_0
= &
{ s_{ag}^2 \spa{b}.c  t_{abc}^8  (
\spb{d}.e \spa{e}.f  [f| K_{abc} | g \ra [a | K_{abc} | d \ra
-\spa{d}.e \spb{e}.f [d| K_{abc} | g \ra [a | K_{abc} | f \ra )
\over
2 \bar N(a,b,c)
N(d,e,f,g)
[c | K_{abc} | g \ra [b | K_{abc} | g \ra
[a | K_{abc} | d \ra  [a | K_{abc} | e \ra  [a | K_{abc} | f \ra  },
\cr}
\equn$$
and,
$$
\eqalign{
H_0
=&
{
s_{ag}^2  s_{bc}^7 t_{def}^7
\left(
\spa{d}.e \spb{e}.f\la f |    K_{abc} | g ] \la a | K_{abc} | d ]
-\spb{d}.e \spa{e}.f \la d | K_{abc} | g ] \la a | K_{abc} | f ]
\right)
\over
2 \spb{b}.c^2
 N(d,e,f)
 { \prod \atop { j=b,c }}  [j | K_{abc} |a  \ra [ j | K_{bc} K_{def} | g ]
{  \prod\atop {i=d,e,f}  }
\la i   | K_{def} K_{bc}  | a \ra  [g | K_{abc} | i \ra
}.
\cr}
\equn$$
Here,
$N(a,b,\cdots m) =\prod_{i<j, i,j\in \{a,b,\cdots m\} } \spa{i}.j$
and
$\bar N(a,b,\cdots m) =\prod_{i<j, i,j\in \{a,b,\cdots m\} } \spb{i}.j$

\subsection{Two Mass Easy Boxes}

The two mass easy boxes have two opposite $\overline{\it MHV}$
three-point corners, an {\it MHV} four-point corner and a
$\overline{\it MHV}$ five-point corner. Again, the terms are relatively
simple and related. They are,

$$
\eqalign{
c[ \{a^-,b^-,c^-\}, d^+, \{e^+,f^+\}, g^+ ] \equiv & W_0,
\cr
c[ \{a^-,b^-,c^+\}, d^+, \{e^+,f^-\}, g^+ ] \equiv & W_1 =
\left(
{
[ c | K_{abc} | f \ra
\over
t_{abc} }
\right)^8 W_0,
\cr
c[ \{a^-,b^-,c^+\}, d^+, \{e^+,f^+\}, g^- ] \equiv & W_2 =
\left(
{
[ c | K_{abc} | g \ra
\over
t_{abc} }
\right)^8 W_0,
\cr
c[ \{a^-,b^+,c^+\}, d^+, \{e^-,f^-\}, g^+ ] \equiv & W_3 =
\left(
{
\spa{e}.f\spb{b}.c
\over
t_{abc} }
\right)^8 W_0,
\cr
c[ \{a^-,b^+,c^+\}, d^+, \{e^-,f^+\}, g^- ] \equiv & W_4 =
\left(
{
\spa{e}.g\spb{b}.c
\over
t_{abc} }
\right)^8 W_0,
\cr
c[ \{a^-,b^+,c^+\}, d^-, \{e^+,f^+\}, g^- ] \equiv & W_5 =
\left(
{
\spa{d}.g\spb{b}.c
\over
t_{abc} }
\right)^8 W_0\cr
c[ \{a^+,b^+,c^+\}, d^-, \{e^-,f^-\}, g^+ ] = & 0,
\cr
c[ \{a^+,b^+,c^+\}, d^-, \{e^-,f^+\}, g^- ] = & 0,
\cr}
\equn$$
where,
$$
W_0 \equiv
{\left(  [g | K_{abc} | d \ra  [d | K_{abc} | g \ra  \right)^2
 \spb{e}.f^2 (t_{abc})^7
(  \la g | K_{abc} k_a k_b k_c | d \ra -  \la g | K_{abc} k_c k_b k_a | d \ra ) \over 2 \spb{a}.b \spb{a}.c \spb{b}.c
\prod_{x=e,f}\spa{x}.d\spa{x}.g \prod_{x=a,b,c} [x | K_{abc} | g \ra [x | K_{abc} | d \ra s_{ef} }.
\equn$$

\subsection{One Mass Boxes}

The one mass boxes have three three-point corners and one massive
six-point corner. For each external helicity configuration there are
(one or) two internal helicity configurations which cause the
massive corner to be either {\it MHV} or {\it NMHV}. These give rise to the
$F_i$ and $P_i$ terms, respectively. Taking the second external
configuration below as an example, $F_1$ and $P_1$ come from,
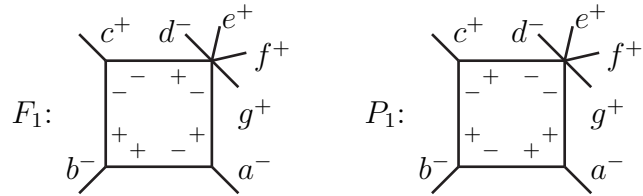
\begin{center}\qquad\qquad
\begin{picture}(110,80)(0,0)
\SetWidth{1}
\Line(10,10)(10,50)
\Line(10,50)(50,50)
\Line(50,50)(50,10)
\Line(50,10)(10,10)
\Line(10,10)(0,0)
\Line(10,50)(0,60)
\Line(50,50)(40,60)
\Line(50,50)(53,63)
\Line(50,50)(63,53)
\Line(50,50)(60,40)
\Line(50,10)(60,0)
\Text(60,10)[l]{$ a^-$}
\Text(-5,10)[l]{$ b^-$}
\Text(8,62)[l]{$ c^+$}
\Text(30,62)[l]{$ d^-$}
\Text(54,67)[l]{$ e^+$}
\Text(66,52)[l]{$ f^+$}
\Text(60,30)[l]{$ g^+$}
\Text(-10,30)[r]{$ F_1$:}
\Text(48,38)[r]{{\scriptsize $-$}}
\Text(48,22)[r]{{\scriptsize $+$}}
\Text(41,16)[r]{{\scriptsize $-$}}
\Text(19,16)[l]{{\scriptsize $+$}}
\Text(12,22)[l]{{\scriptsize $+$}}
\Text(12,38)[l]{{\scriptsize $-$}}
\Text(19,44)[l]{{\scriptsize $-$}}
\Text(41,44)[r]{{\scriptsize $+$}}
\end{picture}\qquad
\begin{picture}(110,80)(0,0)
\SetWidth{1}
\Line(10,10)(10,50)
\Line(10,50)(50,50)
\Line(50,50)(50,10)
\Line(50,10)(10,10)
\Line(10,10)(0,0)
\Line(10,50)(0,60)
\Line(50,50)(40,60)
\Line(50,50)(53,63)
\Line(50,50)(63,53)
\Line(50,50)(60,40)
\Line(50,10)(60,0)
\Text(60,10)[l]{$ a^-$}
\Text(-5,10)[l]{$ b^-$}
\Text(8,62)[l]{$ c^+$}
\Text(30,62)[l]{$ d^-$}
\Text(54,67)[l]{$ e^+$}
\Text(66,52)[l]{$ f^+$}
\Text(60,30)[l]{$ g^+$}
\Text(-10,30)[r]{$ P_1$:}
\Text(48,38)[r]{{\scriptsize $-$}}
\Text(48,22)[r]{{\scriptsize $+$}}
\Text(41,16)[r]{{\scriptsize $+$}}
\Text(19,16)[l]{{\scriptsize $-$}}
\Text(12,22)[l]{{\scriptsize $+$}}
\Text(12,38)[l]{{\scriptsize $-$}}
\Text(19,44)[l]{{\scriptsize $+$}}
\Text(41,44)[r]{{\scriptsize $-$}}
\end{picture}
\end{center}
The $F_i$ terms have the same simplifications as noted above, while
the calculational approach for the resulting $P_i$ terms is
discussed below.
$$
\hspace{1cm}\eqalign{
c[a^-,b^-,c^-,\{d^+,e^+,f^+,g^+\}]=& F_0+0,
\cr
c[a^-,b^-,c^+,\{d^-,e^+,f^+,g^+\}]=& F_1\; + P_1  =
\left( { [ c | K_{abc} | d \ra \over t_{abc} } \right)^8
F_0\; +P_1,
\cr
c[a^-,b^+,c^-,\{d^-,e^+,f^+,g^+\}]=&  F_2\;  + P_2 =
\left( { \spb{b}.c \spa{d}.b \over \spb{a}.c \spa{a}.b}
\right)^8
F_0\; + P_2,
\cr
c[a^-,b^+,c^+,\{d^-,e^-,f^+,g^+\}]=&   F_3\;  + P_3 =
\left( { \spa{d}.e \spb{b}.c  \ra \over t_{abc} } \right)^8
F_0\; + P_3,
\cr
c[a^+,b^-,c^+,\{d^-,e^-,f^+,g^+\}]=& F_4\; +  P_4 =
\left( { \spa{d}.e \spb{a}.c  \ra \over t_{abc} } \right)^8
F_0\; + P_4,
\cr
c[a^+,b^+,c^+  ,\{d^-,e^-,f^-,g^+\}]=& 0 + P_5,
\cr}
\equn$$
where,
$$
F_0
={t_{abc}^6 \spa{a}.b^2 \spa{c}.b^2 \spb{a}.g \spb{d}.e [f| K_{de} K_{abc}|a ]
\over
4 \spb{a}.c \spa{d}.e\spa{e}.f\spa{d}.f \spa{f}.g
[c|K_{abc} |g \ra
{ \prod_{x=d,e,g}} [a|K_{abc} |x \ra }
+{\rm Perm}(d,e,f,g)\,.
\equn
$$

For $P_1$ we use the form of the {\it NMHV} six-point tree amplitude
in appendix
\ref{sixpointmultiplet}. 
We then get,
 $$
 P_1^{[a,b,c,d,e,f,g]} =\left( {\cal T}^1_1 \right)
 +\left( {\cal T}^2_1 + {\cal T}^3_1 + {\cal T}^4_1 + {\cal T}^5_1 \right)|_{\{ (efg)+(feg)+(gef)\} },\equn$$
 with,
 $$\hspace{-0.5cm}
 \eqalign{
 {\cal T}_1^1 &= M_0[a,b,c,d,e,f,g]
 \cr
 \equiv &
 {
 s_{ab}^2 \spa{c}.{d}
 \spa{a}.b^6 \spb{b}.c^2 t_{efg}^7
 \Bigl(
 [ d | K_{efg} | e \ra \spb{e}.f \spa{f}.g  [ g | K_{abc} | c \ra
 -[ d | K_{efg} | g \ra \spb{g}.f \spa{f}.e [ e | K_{abc} | c \ra   \Bigr)
 \over
 2 t_{abc}^2  \spa{a}.{c} [d | K_{abc} | a  \ra \spa{e}.f \spa{f}.g \spa{e}.g
 \prod_{x=e,f,g} \la c | K_{abc}  K_{efg} | x \ra [d | K_{efg} | x \ra
 },
 \cr
  {\cal T}^2_1
 &= \left({ [ e | K_{dfg} | d \ra \over t_{dfg} } \right)^8 M_0[a,b,c,e,d,f,g],
 \cr}
$$
$$\hspace{-0.5cm}
\eqalign{
 {\cal T}^3_1
 & =
 {
 {
  s_{ab}^2 \spa{a}.b^6 \spb{b}.c^2\spb{f}.g^8  \la c |K_{abc}| e ]
 \Bigl(
 \la e | K_{dfg}| d] \spa{d}.f\spb{f}.g \spa{g}.{c}
-\la e | K_{dfg}| g] \spa{g}.f\spb{f}.d \spa{d}.{c}
 \Bigr)
 \over
  2 t_{dfg} \spa{a}.c\spa{a}.e
  \spb{d}.f \spb{f}.g \spb{d}.g
 \prod_{x=d,f,g} [x  | K_{efg} | c \ra [x | K_{efg} | e \ra
 }
 },
 \cr
 {\cal T}^4_1
 & =
 { - [ e | K_{fg} | c \ra^7  \spb{b}.c^2 \spa{a}.b^6  s_{de}s_{fg} s_{ab}^2
 \over
 2 \la c | K_{fg}K_{ade}| a \ra
 \spa{a}.c   \spb{d}.e^2 { \prod \atop {y=d,e} }\la c | K_{abc} | y]
 \spa{f}.g^2 \spa{g}.{c}\spa{f}.{c}
 { \prod \atop {x=f,g} }
 \la c | K_{abc} K_{de} | x \ra
 [d | K_{fg} | c \ra
 },
 \cr
 {\cal T}^5_1 &=
 {\spa{a}.b^6\spb{b}.c^2\spa{c}.d^7\spb{f}.g^6 s_{de}s_{fg}t_{abc} s_{ab}^2
 \over
 2 \spa{a}.{c}
   \la c | K_{abfg}K_{afg} | a \ra  \spa{d}.e^2 \spa{c}.e
 { \prod_{y=f,g} }[ y | K_{abc} | c \ra  [ y    | K_{de} | c \ra
 \prod_{x=d,e} \la x   | K_{fg}K_{abc} | c \ra
 }.
 \cr}
 \equn
 $$
$P_1$ and $P_2$ are related by,
$$
P_2^{[a,b,c,d,e,f,g]}=\left(\frac{\kspa ca}{\kspa bc}\right)^8
P_1^{[a,b,c,d,e,f,g]}.\equn
$$

$P_3$ is obtained by using the {\it NMHV} amplitude of Cachazo and Svr\v cek
\cite{CSgravity}. We then get,
$$
P_3^{[a,b,c,d,e,f,g]}=\sum_{i=1}^{13}{\cal T}_3^i,\equn
$$
where,
$$\hspace{-1.4cm}
\eqalign{
    {\cal T}_3^1=&
        \frac{\kspb ab^2\kspb bc^2\kspa de\kspab a{de}f^7\big(\kspab a{de}f
        \kspab g{ef}d\kspab c{ab}g-\kspab c{ab}d\kspb fg\kspa ga
        \kinv{def}\big)}{2\kspa ag\kspb de\kspb ef^2
        \kspa gc \kinv{def}\kspab a{ef}d{\prod\atop{x=c,g}}
        {\prod\atop{y=d,f}}\kspab x{def}y},\cr
}$$ $$\hspace{-0.5cm}\eqalign{
    {\cal T}_3^2=&
        \frac{\kspb ab^2\kspb bc^2\kspa ea
        \big(\kspa da\kspab c{ab}f+\kspa ca\kspa de\kspb ef\big)^7}
      {2\kspa ca\kspa dg\kspa dc
        \kspab c{ab}e\kspb ef^2\kspa gc
       \kspab c{dg}f}\cr
      &\times\frac{
        \big(\kspa da\kspab c{ab}f+\kspa ca\kspa de\kspb ef\big)
        \kspaa c{ab}{ef}g\kspb gd-\kspab c{ab}d\kspb fg\kspa gd
        \kspaa a{ef}{dg}c}{ \kspaa a{ef}{dg}c\kspaa
        c{ab}{ef}d{\prod\atop{
        x=c,g}}
        \kspaa c{ab}{ef}x\big(\kspa ga\kspab c{ab}f+
        \kspa ca\kspa ge\kspb ef\big)},\cr
}$$ $$\hspace{-0.7cm}\eqalign{
    {\cal T}_3^3=&
        \frac{\kspb ab^2\kspb bc^2
          \kspb gf\kspaa e{gf}{bc}a^7}{2\kspab a{bc}d\kinv{abc}\kspa fg
        \kspa fe^2
        \kspab c{ab}d \kinv{gfe}\kspaa g{fe}{bc}a}\cr
      &\hspace{5cm}\times\frac{\kspaa e{gf}{bc}a
        \kspab g{fe}d\kspa dc+\kspa cg\kspa ed\kspab a{bc}d
        \kinv{gfe}}{{\prod\atop{x=e,g}}\kspab x{efg}d\kspaa c{ab}{efg}x},\cr
}$$ $$\hspace{-6.5cm}\eqalign{
    {\cal T}_3^4=&
        \frac{\kspb ab^2\kspb bc^2\kspab a{bc}f
        \big(\kspa ce\kspab a{bc}g+\kspa ca\kspa ef\kspb fg\big)^7
        }{2\kspa ca\kspb gd
        \kspa cf\kspa fe^2{\prod\atop{x=d,g}}\kspab c{ab}x
        \kspab c{fe}x
        }\cr
}$$ $$\hspace{1.5cm}\eqalign{
      &\times\frac{\big(\kspa ce\kspab a{bc}g+\kspa ca\kspa
        ef\kspb fg\big)
        \kspab c{fe}d\kspa dg+\kspa gc\kspa ed\kspb dg
        \kspaa a{dg}{ef}c}{\kspaa a{dg}{ef}c\kspaa c{ab}{fe}c
        \kspaa c{ab}{dg}e\big(\kspa ce\kspab a{bc}d+\kspa ca\kspa
      ef\kspb fd\big)},\cr
}$$ $$\hspace{-7.4cm}\eqalign{
    {\cal T}_3^5=&
        \frac{\kspb ab^2\kspb bc^2\kspa ae^7\kspa dg\kspb fg^7\kinv{abc}}{2
        \kspb df\kspb dg\kspa ec \kinv{dfg}{\prod\atop{x=a,c}}\kspab x{dg}f
        {\prod\atop{y=d,g}}\kspab e{dfg}y},\cr
}$$ $$\hspace{-1.1cm}\eqalign{
    {\cal T}_3^6=&
        \frac{\kspb ab^2\kspb bc^2
          \kspa ca^7\kspa de^7\kspa ga\kspb fg^7\kspab a{bc}d}{2\kspa dc
        {\prod\atop{x=f,g}}\kspab c{ab}x\kspa ec\kspaa a{fg}{de}c
        \kspab c{de}f
        \kspaa c{ab}{fg}e}\cr
      &\hspace{3cm}\times\frac1{\big(\kspa da\kspab c{ab}f+\kspa ca\kspa dg
        \kspb gf\big)\big(\kspa ce\kspab a{bc}g+\kspa ca\kspa ed
        \kspb dg\big)},\cr
}$$ $$\hspace{-1.4cm}\eqalign{
    {\cal T}_3^7=&
        \frac{\kspb ab^2\kspb bc^2
          \kspa ae^7\kspa cd\kspab a{bc}f^7\kspab c{ab}g}
        {2\kspa ca\kspa ag
        \kspb df\kspab a{bc}d\kspa eg\kspaa a{eg}{df}c\kspab a{eg}f
        \kspaa e{df}{bc}a
        }\cr
      &\hspace{3cm}\times\frac1{\big(\kspa cg\kspab a{bc}f+\kspa ca\kspa gd
        \kspb df\big)\big(\kspa ea\kspab c{ab}d+\kspa ca\kspa eg
        \kspb gd\big)},\cr
 }$$ $$\hspace{-5.2cm}\eqalign{
   {\cal T}_3^8=&
        \frac{\kspb ab^2\kspb bc^2
          \kspa de^7\kspab a{bc}f^7\kspb dg}{2\kspa dg
        \kspab c{ab}f\kinv{abc}\kspa ge \kinv{deg}{\prod\atop{x=d,g}}
        \kspab x{deg}f{\prod\atop{y=a,c}}\kspaa y{abc}{dg}e},\cr
}$$\newpage $$\hspace{-1.3cm}\eqalign{
    {\cal T}_3^9=&
        \frac{\kspb ab^2\kspb bc^2
          \kspa ae^8\kspab c{ab}f\kspab a{bc}g^7}
        {2\kspa ca\kspa af\kspb dg\kspab a{bc}d\kspa ef^2
        \kspaa a{ef}{dg}c
        \kspaa a{ef}{bc}a
        \kspaa e{dg}{bc}a}\cr
      &\hspace{4cm}\times\frac{\kspa de
        \kspa gc\kspab a{bc}d\kspab a{ef}g-\kspa eg\kspab a{bc}g
        \kspa cd \kspab a{ef}d}{
        {\prod\atop{x=d,g}}\kspab a{ef}x
        \big(\kspa ea\kspab c{ab}x+\kspa ca\kspa ef\kspb fx\big)},\cr
}$$ $$\hspace{-2.3cm}\eqalign{
    {\cal T}_3^{10}=&
        \frac{\kspb ab^2\kspb bc^2
          \kspa de^8\kspb df\kspab a{bc}g^7}
        {2\kspa df\kspab c{ab}g\kinv{abc}\kspa ef^2\kinv{def}}
      \frac{\kspa ae\kspa cg
        \kinv{abc}\kspab d{ef}g+\kspa eg\kspab a{bc}g
        \kspaa c{ab}{ef}d}{{\prod\atop{x=d,e}}{\prod\atop{y=a,c}}
        \kspab x{def}g\kspaa y{abc}{def}x},\cr
}$$ $$\hspace{-2.3cm}\eqalign{
    {\cal T}_3^{11}=&
        \frac{\kspb ab^2\kspb bc^2
          \kspab a{bc}f^8\kspa ce\kspa da^7}
        {2\kspa ca\kspab a{bc}e\kspa gd\kspa ga\kspb fe^2
        \kspaa a{dg}{ef}c
       \kspaa a{fe}{bc}a
        \kspab a{dg}f}\cr
      &\hspace{2.1cm}\times\frac{\kspb fg
        \kspab c{ab}d\kspa ag\kspaa d{fe}{bc}a+\kspb fd
        \kspa da\kspab c{ab}g \kspaa g{fe}{bc}a}{
        {\prod\atop{x=d,g}}\kspaa x{fe}{bc}a
        \big(\kspa cx\kspab a{bc}f+\kspa ca\kspa xe\kspb ef\big)},\cr
}$$ $$\hspace{-2.2cm}\eqalign{
    {\cal T}_3^{12}=&
        \frac{\kspb ab^2\kspb bc^2
          \kspb gf^8\kspa ge\kspa da^7\big(-\kspab a{bc}f
        \kspab c{ab}d\kspab d{fe}g+\kspb fd\kspa da\kinv{abc}
        \kspab c{fe}g
        \big)}{2\kspb ge\kspa cd\kspb fe^2\kinv{gfe}
        {\prod\atop{x=a,c,d}}{\prod\atop{y=f,g}}\kspab x{efg}y},\cr
 }$$ $$\hspace{-1.7cm}\eqalign{
   {\cal T}_3^{13}=&
        \frac{\kspb ab^2\kspb bc^2\kspa ad\kspab a{bc}g
        \big(\kspa ea\kspab c{ab}f+\kspa ca\kspa ed\kspb df\big)^8}
        {2\kspa ca{\prod\atop{x=d,f}}\kspab c{ab}x\kspb df\kspa eg\kspa ce
        \kspa gc\kspaa a{df}{eg}c
        \kspab c{eg}f\kspaa c{ab}{df}e}\cr
      &\hspace{3.3cm}\times\frac1{\big(\kspa ga\kspab c{ab}f+\kspa ca\kspa gd
        \kspb df\big)\big(\kspa ce\kspab a{bc}d+\kspa ca\kspa eg
        \kspb gd\big)}.
}\equn
$$

\hfill

$P_4$ has the additional complication that we must sum over the full
$\mathcal N=8$ multiplet running in the loop.
We obtain a form based on $P_3$
with relative factors for
each ${\cal T}_3^i$.
We also obtain one
extra term which is not present in $P_3$. We get,
$$
P_4^{[a,b,c,d,e,f,g]}=\sum_{i_1}^{13}\big(Y_4^i)^8{\cal T}_3^i
+{\cal T}_4^{14},\equn
$$
where,
$$
\eqalign{
    Y_4^1=&-\frac{\kspab b{de}f}{\kspab a{de}f},\cr
    Y_4^2=&\frac{\kspa bd\kspab c{ab}f+\kspa bc\kspa de\kspb ef}
            {\kspa da\kspab c{ab}f+\kspa ca\kspa de\kspb ef},\cr
    Y_4^3=&-\frac{\kspaa b{ac}{fg}e}{\kspaa a{bc}{fg}e},\cr
    Y_4^4=&-\frac{\kspa ce\kspab b{ac}g+\kspa cb\kspa ef\kspb fg}
            {\kspa ce\kspab a{bc}g+\kspa ca\kspa ef\kspb fg},\cr
    Y_4^5=&\frac{\kspa eb}{\kspa ae},\cr
    Y_4^6=&\frac{\kspa bc}{\kspa ca},\cr
    Y_4^7=&\frac{\kspa eb\kspab a{bc}f-\kspa ba\kspa ed\kspb df}
            {\kspa ae\kspab a{bc}f},\cr
    Y_4^8=&-\frac{\kspab b{ac}f}{\kspab a{bc}f},\cr
    Y_4^9=&\frac{\kspa be\kspab a{bc}g+\kspa ba\kspa ed\kspb dg}
            {\kspa ea\kspab a{bc}g},\cr
    Y_4^{10}=&-\frac{\kspab b{ac}g}{\kspab a{bc}g},\cr
    Y_4^{11}=&\frac{\kspa bd\kspab a{bc}f+\kspa ba\kspa de\kspb ef}
            {\kspa da\kspab a{bc}f},\cr
    Y_4^{12}=&\frac{\kspa db}{\kspa ad},\cr
    Y_4^{13}=&\frac{\kspa be\kspab c{ab}f+\kspa bc\kspa ed\kspb df}
            {\kspa ea\kspab c{ab}f+\kspa ca\kspa ed\kspb df},\cr
  {\cal T}_4^{14}=&\left(\frac{\kspa ab}{\kspa ca}\right)^8{\cal T}_3^6
(a\leftrightarrow c).
}\equn
$$

Last comes $P_5$ which has been obtained from the amplitude of Cachazo
and Svr\v cek by
letting 5 and 6 be the internal gravitons. We get,
$$\hspace{1cm}
\eqalign{
P_5^{[a,b,c,d,e,f,g]}=&{\cal T}_5^1+
{\cal T}_5^1(d\leftrightarrow e)
        +{\cal T}_5^2+{\cal T}_5^2(a\leftrightarrow c)\cr
        &\hspace{-0.4cm}+{\cal T}_5^3+{\cal T}_5^3(d\leftrightarrow e)
        +{\cal T}_5^3(a\leftrightarrow c)+{\cal T}_3^c(a\leftrightarrow c,
        d\leftrightarrow e)\cr
        &\hspace{-0.4cm}+{\cal T}_5^4+{\cal T}_5^4(d\leftrightarrow e)
        +{\cal T}_5^5+{\cal T}_5^5(a\leftrightarrow c)+{\cal T}_5^6,
}\equn
$$
where,
$$
\eqalign{
    {\cal T}_5^1=&
        \frac{\kspb ab^2\kspb bc^2\kspa ef\kspab d{ef}g^7\big(\kspab d{ef}g
        \kspab a{fg}e\kspab c{ab}d-\kspb de\kspab c{ab}g\kspa ad
        \kinv{efg}\big)}{2\kspa da\kspa cd\kspb ef\kspb fg^2
        \kinv{efg}\kspab d{fg}e{\prod\atop{x=e,g}}
        \kspab a{efg}x\kspab c{efg}x},\cr
    {\cal T}_5^2=&
        \frac{\kspb ab^2\kspb bc^2
          \kspab c{ab}g\kspaa f{de}{bc}a^7}{2\kspa ca
        \kspa ag\kspa gf^2
        \kspb ed\kspaa a{gf}{de}c\kspaa a{fg}{bc}a}\cr
      &\hspace{2.5cm}\times\frac{\kspaa f{de}{bc}a
        \kspab a{gf}e\kspa ec-\kspa fe\kspab a{bc}e
        \kspaa a{gf}{de}c}{{\prod\atop{x=d,e}}\kspab a{bc}x
        \kspab a{fg}x
        \big(\kspa fa\kspab c{ab}x+\kspa ca\kspa fg\kspb gx\big)},\cr
    {\cal T}_5^3=&
        \frac{\kspb ab^2\kspb bc^2
          \kspa df^7\kspa ea\kspab c{ab}g^7\kspab a{bc}d}
        {2\kspa ca\kspa dc
        \kspb eg\kspab c{ab}e\kspa fc\kspaa a{eg}{df}c
        \kspab c{df}g\kspaa f{eg}{ab}c
        }\cr
      &\hspace{2.5cm}\times\frac1{\big(\kspa da\kspab c{ab}g+\kspa ca\kspa de
        \kspb eg\big)\big(\kspa cf\kspab a{bc}e+\kspa ca\kspa fd
        \kspb de\big)},\cr
    {\cal T}_5^4=&
        \frac{\kspb ab^2\kspb bc^2
          \kspa df^8\kspb dg\kinv{abc}^7\big(\kspa ef
        \kspab a{bc}e\kspaa c{ab}{fg}d+\kspa fa\kinv{abc}\kspa ce
        \kspab d{fg}e\big)}{2\kspa gd
        \kspa fg^2\kinv{dfg}
        {\prod\atop{x=d,f}}{\prod\atop{y=a,c}}\kspab x{dfg}e
        \kspab y{abc}e\kspaa y{abc}{dfg}x},\cr
    {\cal T}_5^5=&
        \frac{\kspb ab^2\kspb bc^2
          \kspab a{bc}g^8\kspa cf\kspa ed^7}{2\kspa ca\kspab a{bc}f
          \kspa ae\kspa ad\kspb gf^2
        \kspaa a{de}{gf}c
        \kspab a{de}g
        }\cr
      &\hspace{2.5cm}\times\frac{\kspab c{ab}g\kspb ed
        \kspa da\kspaa e{fg}{bc}a+\kspb ge\kspa de\kspab c{ab}d
        \kspaa a{fg}{bc}a}{\kspaa a{fg}{bc}a
        {\prod\atop{x=d,e}}
        \big(\kspa cx\kspab a{bc}g+\kspa ca\kspa xf\kspb fg\big)
      \kspaa x{fg}{bc}a},\cr
    {\cal T}_5^6=&
        \frac{\kspb ab^2\kspb bc^2
          \kspa de\kinv{abc}\kspab f{de}g^8}{2\kspb de
        \kspb dg\kspb eg\kspa fa\kspa fc\kinv{deg}
        \kspab a{de}g\kspab c{de}g\kspab f{eg}d\kspab f{dg}e}.
}\equn
$$
\

\section{Relations Between  Box Coefficients}

The box-coefficients exhibit a large number of relations. As a
consequence of the IR structure many combinations can be used to
create expressions for the tree amplitudes. This has in fact been
used to obtain relatively compact formulae for tree
amplitudes~\cite{BDKn,Roiban:2004ix} and gave rise to the BCFW
recursion relations~\cite{Britto:new}. Since the IR relations are
satisfied, the box-coefficients are related to the tree amplitudes
and in fact yield a form of the tree amplitude which is equivalent
to that obtained via recursion.  Before commencing it is convenient
to define scaled box-coefficients\footnote{The scaling factors are
essentially the momentum prefactors appearing in the integral
functions.},
$$
\eqalign{
\hat c^{1m}[a,b,c,\{d,e,f,g\}] &\equiv
 { c^{1m}[a,b,c,\{d,e,f,g\}] \over s_{ab} s_{bc} },
\cr
\hat c^{2m\; h}[a,\{b,c\},\{d,e,f\},g] &\equiv
{ c^{2m\; h}[a,\{b,c\},\{d,e,f\},g] \over s_{ga}t_{abc} },
\cr
\hat c^{2m\; e}[a,\{b,c\},d,\{e,f,g\}]
&\equiv
{  c^{2m\; e}[a,\{b,c\},d,\{e,f,g\}] \over
t_{abc}t_{bcd}-s_{bc}t_{efg} }.
\cr}\equn
$$
We will also use this notation to indicate the scaled functions
which define the box-coefficients.

\subsection{Expressions for tree amplitudes}

For the seven-point one-loop {\it NMHV} amplitude there are circa 1000
independent boxes with each box coefficient containing two or more
terms.  We can extract the tree by looking at the coefficient of a
specific logarithm: there being three independent choices:
$\ln(-s_{12})$, $\ln(-s_{45})$ and $\ln(-s_{34})$. If we take the
coefficient of $\ln(-s_{12})$ then only a subset of boxes will
contribute to this. Contained within this is a further subset where
the legs $1$ and $2$ are massless and the boxes are the one-mass and
two-mass hard.
$$
\eqalign{
 M_7^{\rm tree}
 = &
\left(
 \hat F_0^{[1,2,3,4,5,6,7]}+ \{ 1 \leftrightarrow 2 \}
\right)
\cr+&
\left(
\hat F_1^{[1,2,4,3,5,6,7]}+\hat F_1^{[1,2,5,3,4,6,7]}+\hat F_1^{[1,2,6,3,5,4,7]}+\hat F_1^{[1,2,7,3,5,6,4]}
+ \{ 1 \leftrightarrow 2 \}
\right)
\cr+&
\left( \hat P_1^{[1,2,3,4,5,6,7]}+\hat P_1^{[1,2,3,5,4,6,7]}+\hat P_1^{[1,2,3,6,5,4,7]}+\hat P_1^{[1,2,3,7,5,6,4]}
+ \{ 1 \leftrightarrow 2 \} \right)
\cr+&
\left(  \hat G_8^{[2,3,4,5,6,7,1]}+\hat G_8^{[2,3,5,4,6,7,1]}+\hat G_8^{[2,3,6,5,4,7,1]}+\hat G_8^{[2,3,7,5,6,4,1]}
+ \{ 1 \leftrightarrow 2 \} \right)
\cr+&
\left( \hat H_8^{[2,3,4,5,6,7,1]}+\hat H_8^{[2,3,5,4,6,7,1]}+\hat H_8^{[2,3,6,5,4,7,1]}+\hat H_8^{[2,3,7,5,6,4,1]}
+\{ 1 \leftrightarrow 2 \} \right)
\cr+&
\Bigl( \hat G_9^{[2,4,5,3,6,7,1]}+\hat G_9^{[2,4,6,3,5,7,1]}+\hat G_9^{[2,4,7,3,6,5,1]}+
               \hat G_9^{[2,5,6,3,4,7,1]}
\cr &\hskip 4.0truecm
+\hat G_9^{[2,5,7,3,4,6,1]}+\hat G_9^{[2,6,7,3,4,5,1]}+ \{ 1 \leftrightarrow 2 \} \Bigr)
\cr+&
\Bigl(  \hat H_9^{[2,4,5,3,6,7,1]}+\hat H_9^{[2,4,6,3,5,7,1]}+\hat H_9^{[2,4,7,3,6,5,1]}+
               \hat H_9^{[2,5,6,3,4,7,1]}
\cr &\hskip 4.0truecm
+\hat H_9^{[2,5,7,3,4,6,1]}+\hat H_9^{[2,6,7,3,4,5,1]} + \{ 1 \leftrightarrow 2 \}  \Bigr).
\cr}
\equn$$
Within this set there are two subsets which each yield the tree, e.g.
$$\hspace{-0.1cm}
\eqalign{
\hat F_0^{[1, 2, 3, 4, 5, 6, 7]} +
 \biggl(\sum_{(a,b,c,d) \in  {\cal S}_1 }\!\!\!\!\!\!\hat F_1^{[1, 2, a, 3, b, c, d]}\biggr)
+\biggl(\sum_{(a,b,c,d) \in  {\cal S}_1 }\!\!\!\!\!\!\hat G_8^{[1, 3, a, b, c, d, 2]}\biggr)
+ \biggl( \sum_{(a,b,c,d) \in  {\cal S}_2 }\!\!\!\!\!\!
\hat G_9^{[1, a, b, 3, c, d, 2]}\biggr)   
\cr
+ \biggl(  \sum_{(a,b,c,d) \in  {\cal S}_1 }
\!\!\!\!\!\!\hat H_8^{[2, 3, a, b, c, d, 1]}\biggr)
+ \biggl(  \sum_{(a,b,c,d) \in  {\cal S}_2 } \!\!\!\!\!\!\hat H_9^{[2, a, b, 3, c, d, 1]}\biggr) 
+  \biggl( \sum_{(a,b,c,d) \in  {\cal S}_1  }\!\!\!\!\!\!\hat P_1^{[2, 1, 3, a, b, c, d]}\biggr),
\cr}
\equn$$
where,
$$\hspace{0.5cm}
\eqalign{ {\cal S}_1 &=  \{ (4,5,6,7), (5,4,6,7),(6,4,5,7),
(7,4,5,6) \}, \cr {\cal S}_2 &=  \{
(4,5,6,7),(4,6,5,7),(4,7,5,6),(5,6,4,7),(5,7,4,6),(6,7,4,5) \}. \cr}
\equn$$
This provides a fairly compact realisation of the
seven-point tree amplitude containing twenty-nine individual terms.
This collection of terms corresponds exactly to the terms that
would be obtained from a recursive calculation using legs $1$ and
$2$ for the recursion. The above expression has all the necessary
symmetries although not all are manifest.

This subset of the box-coefficients corresponds to those terms where 
legs $1$ and $2$ are isolated at massless corners and where these corners have the
helicity structure indicated.
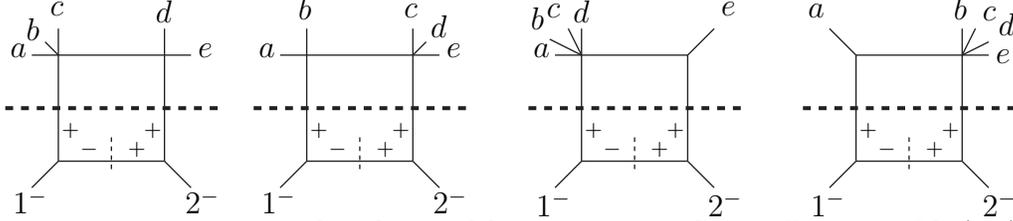
\vspace{-0.9cm}
\begin{center}\hspace{-2cm}
\begin{picture}(100,110)(-30,0)

\Line(30,30)(30,70)
\Line(70,30)(70,70)
\Line(30,30)(70,30)
\Line(70,70)(30,70)

\Line(30,30)(20,20)
\Line(70,30)(80,20)

\Line(30,70)(20,70)
\Line(30,70)(30,80)
\Line(30,70)(25,75)

\Line(70,70)(70,80)
\Line(70,70)(80,70)

\Text(13,15)[l]{$1^-$}
\Text(78,15)[l]{$2^-$}
\Text(12,72)[l]{$a$}
\Text(18,80)[l]{$b$}
\Text(27,88)[l]{$c$}

\Text(67,87)[l]{$d$}
\Text(83,72)[l]{$e$}

\Text(35,40)[c]{$^+$}
\Text(66,40)[c]{$^+$}
\Text(60,33)[c]{$^+$}
\Text(42,33)[c]{$^-$}

\DashLine(50,39)(50,27){2}

\SetWidth{1.5}
\DashLine(10,50)(90,50){3}

\end{picture}
\begin{picture}(100,110)(-20,0)

\Line(30,30)(30,70)
\Line(70,30)(70,70)
\Line(30,30)(70,30)
\Line(70,70)(30,70)

\Line(30,30)(20,20)
\Line(70,30)(80,20)

\Line(30,70)(20,70)
\Line(30,70)(30,80)

\Line(70,70)(70,80)
\Line(70,70)(80,70)
\Line(70,70)(75,75)

\Text(13,15)[l]{$1^-$}
\Text(78,15)[l]{$2^-$}
\Text(12,72)[l]{$a$}
\Text(27,88)[l]{$b$}

\Text(67,87)[l]{$c$}
\Text(77,81)[l]{$d$}
\Text(83,72)[l]{$e$}

\Text(35,40)[c]{$^+$}
%
\Text(66,40)[c]{$^+$}
\Text(60,33)[c]{$^+$}
\Text(42,33)[c]{$^-$}

\DashLine(50,39)(50,27){2}

\SetWidth{1.5}
\DashLine(10,50)(90,50){3}

\end{picture}
\begin{picture}(100,110)(-20,0)

\Line(30,30)(30,70)
\Line(70,30)(70,70)
\Line(30,30)(70,30)
\Line(70,70)(30,70)

\Line(30,30)(20,20)
\Line(70,30)(80,20)

\Line(30,70)(20,70)

\Line(30,70)(25,80)
\Line(30,70)(18,76)

\Line(30,70)(30,80)

\Line(70,70)(80,80)

\Text(13,14)[l]{$1^-$}
\Text(78,14)[l]{$2^-$}

\Text(12,72)[l]{$a$}
\Text(11,84)[l]{$b$}
\Text(17,87)[l]{$c$}
\Text(27,87)[l]{$d$}

\Text(83,88)[l]{$e$}

\Text(35,40)[c]{$^+$}
\Text(66,40)[c]{$^+$}
\Text(60,33)[c]{$^+$}
\Text(42,33)[c]{$^-$}

\DashLine(50,39)(50,27){2}

\SetWidth{1.5}
\DashLine(10,50)(90,50){3}

\end{picture}
\begin{picture}(100,110)(-20,0)

\Line(30,30)(30,70)
\Line(70,30)(70,70)
\Line(30,30)(70,30)
\Line(70,70)(30,70)

\Line(30,30)(20,20)
\Line(70,30)(80,20)

\Line(30,70)(20,80)

\Line(70,70)(70,80)

\Line(70,70)(80,75)
\Line(70,70)(75,80)

\Line(70,70)(80,70)

\Text(13,14)[l]{$1^-$}
\Text(78,14)[l]{$2^-$}
\Text(12,87)[l]{$a$}

\Text(67,88)[l]{$b$}
\Text(78,86)[l]{$c$}
\Text(84,82)[l]{$d$}

\Text(83,70)[l]{$e$}

\Text(35,40)[c]{$^+$}
\Text(66,40)[c]{$^+$}
\Text(60,33)[c]{$^+$}
\Text(42,33)[c]{$^-$}

\DashLine(50,39)(50,27){2}

\SetWidth{1.5}
\DashLine(10,50)(90,50){3}

\end{picture}
\end{center}
\vspace{-0.8cm}
Alternate expressions may be obtained by examining the coefficients of
$\ln(s_{45})$ and $\ln(s_{34})$.

\subsection{The coefficient of $\ln(-t_{123})$ }

A different type of relationship holds for the box-coefficients
which contribute to the soft divergence $\ln(-t_{123})/\eps$. These
soft divergences are absent so the box coefficients are
conspiring to make them cancel. There are three types of box giving
this divergence: two-mass easy, two-mass hard and one-mass.
Specifically we must have,\vspace{-0.1cm}
$$
\eqalign{
& \left(
\sum_{Z_{(1,2,3)}}
\sum_{Z_{(4,5,6,7)}}
 \hat c^{2m h}_{( 1^-, \{ 2^-,3^- \} ,\{ 4^+,5^+,6^+ \}, 7^+ )}
\right)
-
\left(
\sum_{Z_{(1,2,3)}}
\hat c^{1m}_{ (1^-, 2^-,3^- ,\{4^+,5^+,6^+, 7^+\} )}
\right)
\cr
&\hspace{7.15cm}
-\left(
\sum_{P_{(4,5,6,7)}}
\hat c^{2m e }_{
( \{ 1^-, 2^-,3^-\} , 4^+,\{ 5^+,6^+\}, 7^+ )}
\right)
=0,
\cr}
\hspace{-1cm}\equn\label{Log123eq}
$$
where $Z$ denotes cyclic permutations and,
$$
P_{(4,5,6,7)}=\{(4,5,6,7),(4,7,5,6),(4,6,7,5),(5,4,6,7),(5,4,7,6),(6,4,5,7)\}.
$$
This relationship is indeed satisfied since,
$$\hspace{-1cm}
\eqalign{
\sum_{Z_{(1,2,3)}}
\sum_{Z_{(4,5,6,7)}}
 \hat c^{2m h}_{( 1^-, \{ 2^-,3^- \} ,\{ 4^+,5^+,6^+ \}, 7^+ )}
=
2\sum_{Z_{(1,2,3)}}
\hat c^{1m}_{ (1^-, 2^-,3^- ,\{ 4^+,5^+,6^+, 7^+\} )},
\cr
\sum_{P_{(4,5,6,7)}}
\hat c^{2m e }_{(\{ 1^-, 2^-,3^-\} ,4^+,\{ 5^+,6^+\}, 7^+ )}
=
\sum_{Z_{(1,2,3)}}
\hat c^{1m}_{ (1^-, 2^-,3^- ,\{ 4^+,5^+,6^+ \}, 7^+ )},
\cr}
\equn
$$
although clearly these two constraints are considerably stronger than the single constraint~(\ref{Log123eq}).

\section{Six-Point Tree amplitudes involving non-gravitons}
\label{sixpointmultiplet}
\def\spaN(#1,#2){\left\langle#1#2\right\rangle_{\!}}
\def\spbN(#1,#2){\left[#1#2\right]_{\!}}
\def\spabt(#1,\{ #2,#3,#4\},#5){\langle#1\,|P_{#2 #3 #4}|#5]}
\def\spabtt(#1,\{#2,#3,#4\},#5){\langle#1|P_{#2 #3 #4}|#5]}
\def\spbatt(#1,\{#2,#3,#4\},#5){[#1|P\!_{#2 #3 #4}|#5\rangle}
\def\spbattN(#1,\{#2,#3\},#4){[#1|P\!_{#2 #3}|#4\rangle}
\def\spbattt(#1,\{#2,#3\},#4){[#1|P\!_{#2 #3}|#4\rangle}
\def\spaattt(#1,\{#2,#3\},\{#4,#5\},#6){\langle#1|_{\!}P\!_{#2 #3}P\!_{#4 #5}|#6\rangle_{\!}}
\def\spabt(#1,\{#2,#3,#4\},#5){\langle#1|_{\!}P_{\!#2 #3 #4}|#5]_{\!}}
\def\spabtt(#1,\{#2,#3,#4\},#5){\langle#1|_{\!}P_{\!#2 #3 #4}|#5]_{\!}}
\def\spbatt(#1,\{#2,#3,#4\},#5){[#1|_{\!}P_{\!#2 #3 #4}|#5\rangle_{\!}}
To calculate the cuts of the seven-point amplitude we need
the six-point {\it NMHV} amplitudes where one pair of particles is of arbitrary type.
The six-point amplitude is given in the form,
$$
M( 1^-, 2^-,(l_1)^-_h , (l_2)^+_h, 5^+, 6^+)
=\sum_{i=1}^{14} T_i(h) = \sum_{i=1}^{14} A_i (X_i)^{2h},
\equn$$
where $h=2$ for a graviton, $h=3/2$ for a gravitino, $h=1$ for a
vector, $h=1/2$ for a Dirac fermion and $h=0$ for a scalar particle.
The expression is also valid for negative values of $h$ provided we recognise
that this corresponds to a particle of the opposite helicity e.g.
$1^-_{-2} \equiv 1^+_{+2}$.
The explicit forms of the $T_i$ are given by,
$$\hspace{-1.3cm}\eqalign{
T_1&\!=\!\frac{-i \spaN(1,2)^7 \spaN(5,l_2) \spbN(2,l_1)
   \spbN(5,6)^7}{\spaN(1,l_1) \spaN(2,l_1) \spabtt(1,\{1,2,l_1\},5)
   \spabtt(1,\{1,2,l_1\},l_2) \spabtt(2,\{1,2,l_1\},6) \spabtt(l_1,\{1,2,l_1\},6)
   \spbN(5,l_2) \spbN(6,l_2) \t(1,2,l_1)}\Big[ \delta _{h,2}\Big],\cr
T_2&\!=\!
   \frac{-i
   \spaN(2,l_1) \spabtt(1,\{2,6,l_1\},6)^8 \spbN(5,l_2)}{\spaN(1,5) \spaN(1,l_2)
   \spaN(5,l_2) \spabtt(1,\{2,6,l_1\},2) \spabtt(1,\{2,6,l_1\},l_1)
   \spabtt(5,\{2,6,l_1\},6) \spabtt(l_2,\{2,6,l_1\},6) \spbN(2,6) \spbN(2,l_1)
   \spbN(6,l_1) \t(2,6,l_1)}\!\bigg[\!\!\frac{-i \spaN(1,l_2) \spbN(6,l_1)}{\spabtt(1,\{2,6,l_1\},6)}\!\bigg]^{A},\cr
T_3&\!=\!\frac{-i \spaN(1,2)^7 \spaN(5,l_1) \spbN(2,l_2)
   \spbN(5,6)^7}{\spaN(1,l_2) \spaN(2,l_2) \spabtt(1,\{5,6,l_1\},5)
   \spabtt(1,\{5,6,l_1\},l_1) \spabtt(2,\{5,6,l_1\},6) \spabtt(l_2,\{5,6,l_1\},6)
   \spbN(5,l_1) \spbN(6,l_1) \t(5,6,l_1)}\big[\delta _{h,-2}\big],\cr
T_4&\!=\!\frac{i \spaN(1,l_1)^7 \spaN(2,5)
   \spbN(5,6)^7 \spbN(l_1,l_2)}{\spaN(1,l_2) \spaN(l_1,l_2)
   \spabtt(1,\{2,5,6\},2) \spabtt(1,\{2,5,6\},5) \spabtt(l_1,\{2,5,6\},6) \spabtt(l_2,\{2,5,6\},6)
   \spbN(2,5) \spbN(2,6) \t(2,5,6)}\!\left[\frac{i \spaN(1,l_2)}{\spaN(1,l_1)}\right]^{A},\cr
T_5&\!=\!\frac{i \spaN(1,2)^7
   \spaN(l_1,l_2) \spbN(2,5) \spbN(6,l_2)^7}{\spaN(1,5) \spaN(2,5)
   \spabtt(1,\{1,2,5\},l_1) \spabtt(1,\{1,2,5\},l_2) \spabtt(2,\{1,2,5\},6) \spabtt(5,\{1,2,5\},6)
   \spbN(6,l_1) \spbN(l_1,l_2) \t(1,2,5)}\!\left[\frac{i \spbN(6,l_1)}{\spbN(6,l_2)}\right]^{A},\cr
T_6&\!=\!\frac{-i
   \spaN(1,l_1)^7 \spaN(2,l_2) \spbN(5,l_1) \spbN(6,l_2)^7}{\spaN(1,5)
   \spaN(5,l_1) \spabtt(1,\{1,5,l_1\},2) \spabtt(1,\{1,5,l_1\},l_2)
   \spabtt(5,\{1,5,l_1\},6) \spabtt(l_1,\{1,5,l_1\},6) \spbN(2,6) \spbN(2,l_2)
   \t(1,5,l_1)}\!\left[\frac{-i \spabtt(1,\{1,5,l_1\},6)}{\spaN(1,l_1) \spbN(6,l_2)}\right]^{A},\cr
T_7&\!=\!\frac{i
   \spabtt(1,\{1,5,6\},l_2)^7 (\spaN(1,l_2) \spaN(2,l_1) \spabtt(5,\{1,5,6\},l_1)
   \spbN(2,l_2)-\spaN(1,l_1) \spaN(2,l_2) \spabtt(5,\{1,5,6\},l_2) \spbN(2,l_1))
   \spbN(5,6)}{\spaN(1,6)^2 \spaN(5,6) \spabtt(1,\{1,5,6\},2) \spabtt(1,\{1,5,6\},l_1)
   \spabtt(5,\{1,5,6\},2) \spabtt(5,\{1,5,6\},l_1) \spabtt(5,\{1,5,6\},l_2) \spbN(2,l_1)
   \spbN(2,l_2) \spbN(l_1,l_2) \t(1,5,6)}\!\left[\frac{i \spabtt(1,\{1,5,6\},l_1)}{\spabtt(1,\{1,5,6\},l_2)}\right]^{A},\cr
T_8&\!=\!\frac{i\spabtt(1,\{2,5,l_1\},5)^7 (\spaN(1,5)
   \spaN(2,l_1) \spabtt(l_2,\{2,5,l_1\},l_1) \spbN(2,5)- \spaN(1,l_1) \spaN(2,5)
   \spabtt(l_2,\{2,5,l_1\},5) \spbN(2,l_1)) \spbN(6,l_2)}{\spaN(1,6)^2
   \spaN(6,l_2) \spabtt(1,\{2,5,l_1\},2) \spabtt(1,\{2,5,l_1\},l_1)
   \spabtt(l_2,\{2,5,l_1\},2) \spabtt(l_2,\{2,5,l_1\},5)
   \spabtt(l_2,\{2,5,l_1\},l_1) \spbN(2,5) \spbN(2,l_1) \spbN(5,l_1)
   \t(2,5,l_1)}\!\left[\frac{i \spaN(l_2,1)
   \spbN(5,l_1)}{\spabtt(1,\{2,5,l_1\},5)}\right]^{A},\cr
T_9&\!=\!\frac{i
   \spaN(1,l_1)^8 \spbN(5,l_2)^7 (\spaN(1,l_2) \spaN(2,5)
   \spabtt(l_1,\{1,6,l_1\},2) \spbN(5,l_2)-\spaN(1,2) \spaN(5,l_2)
   \spabtt(l_1,\{1,6,l_1\},l_2) \spbN(2,5)) \spbN(6,l_1)}{\spaN(1,6)^2\hskip -5 pt
   \spaN(6,l_1)\hskip -2 pt  \spabtt(1,\{1,6,l_1\},2) \spabtt(1,\{1,6,l_1\},5)
   \spabtt(1,\{1,6,l_1\},l_2) \spabtt(l_1,\{1,6,l_1\},2) \spabtt(l_1,\{1,6,l_1\},5)
   \spabtt(l_1,\{1,6,l_1\},l_2) \spbN(2,5)\hskip -1 pt  \spbN(2,l_2) \t(1,6,l_1)}\hskip -5 pt
\left[\!\hskip -1 pt \frac{i \spabtt(1,\{1,6,l_1\},5)}{\spaN(l_1,1) \spbN(5,l_2)}\!\right]^{A},\cr
T_{10}&\!=\!\frac{i \spaN(1,2)^8 \spbN(2,6)
   \spbN(5,l_2)^7 (\spaN(1,l_2) \spaN(5,l_1) \spabtt(2,\{1,2,6\},l_1)
   \spbN(5,l_2)-\spaN(1,l_1) \spaN(5,l_2) \spabtt(2,\{1,2,6\},l_2)
   \spbN(5,l_1))}{\spaN(1,6)^2 \spaN(2,6) \spabtt(1,\{1,2,6\},5) \spabtt(1,\{1,2,6\},l_1)
   \spabtt(1,\{1,2,6\},l_2) \spabtt(2,\{1,2,6\},5) \spabtt(2,\{1,2,6\},l_1) \spabtt(2,\{1,2,6\},l_2)
   \spbN(5,l_1) \spbN(l_1,l_2) \t(1,2,6)}\!\left[ \hskip -2 pt \frac{i \spbN(5,l_1)}{\spbN(5,l_2)}\right]^{A},\cr
T_{11}&\!=\!\frac{i \spaN(1,5) \spaN(2,l_1)^7
   \spbN(5,6)^8 (\spaN(2,l_2) \spabtt(l_1,\{1,5,6\},5) \spbN(2,l_1)
   \spbN(6,l_2)-\spaN(2,l_1) \spabtt(l_2,\{1,5,6\},5) \spbN(2,l_2)
   \spbN(6,l_1))}{\spaN(2,l_2) \spaN(l_1,l_2) \spabtt(2,\{1,5,6\},5)
   \spabtt(2,\{1,5,6\},6) \spabtt(l_1,\{1,5,6\},5) \spabtt(l_1,\{1,5,6\},6) \spabtt(l_2,\{1,5,6\},5)
   \spabtt(l_2,\{1,5,6\},6) \spbN(1,5) \spbN(1,6)^2 \t(1,5,6)}\!\left[\frac{i \spaN(2,l_2)}{\spaN(2,l_1)}\right]^{A},\cr
T_{12}&\!=
 \hskip -2 pt
\frac{-i\spaN(1,l_2) \spaN(2,l_1)^7
(\spaN(2,5)
   \spabtt(l_1,\{2,5,l_1\},l_2) \spbN(2,l_1) \spbN(5,6)+ \spaN(2,l_1)
   \spabtt(5,\{2,5,l_1\},l_2) \spbN(2,5) \spbN(6,l_1)) \spbN(6,l_2)^8}{\spaN(2,5)
 \hskip -2 pt  \spaN(5,l_1) \hskip -2 pt \spabtt(2,\{2,5,l_1\},6) \spabtt(2,\{2,5,l_1\},l_2)
   \spabtt(5,\{2,5,l_1\},6) \spabtt(5,\{2,5,l_1\},l_2) \spabtt(l_1,\{2,5,l_1\},6)
   \spabtt(l_1,\{2,5,l_1\},l_2)  \hskip -2 pt\spbN(1,l_2)  \hskip -2 pt \spbN(1,6)^2 \hskip -3 pt \t(2,5,l_1)}
 \hskip -5 pt\left[\frac{i \spabtt(2,\{2,5,l_1\},6)}{\spaN(l_1,2)\spbN(6,l_2)}\!\right]^{A},\cr
T_{13}&\!=\!\frac{-i
   \spaN(1,l_1) \spabtt(2,\{1,6,l_1\},6)^7 (\spaN(2,5) \spabtt(l_2,\{1,6,l_1\},l_1)
   \spbN(2,6) \spbN(5,l_2)+\spaN(5,l_2) \spabtt(2,\{1,6,l_1\},l_1) \spbN(2,5)
   \spbN(6,l_2))}{\spaN(2,5) \spaN(2,l_2) \spaN(5,l_2)
   \spabtt(2,\{1,6,l_1\},l_1) \spabtt(5,\{1,6,l_1\},6) \spabtt(5,\{1,6,l_1\},l_1)
   \spabtt(l_2,\{1,6,l_1\},6) \spabtt(l_2,\{1,6,l_1\},l_1) \spbN(1,6)^2
   \spbN(1,l_1) \t(1,6,l_1)}\!\left[\!\frac{i \spaN(l)2,2) \spbN(6,l_1)}{\spabtt(2,\{1,6,l_1\},6)}\!\right]^{A},\cr
T_{14}&\!=\!\frac{i \spaN(1,2)
   \spabtt(l_1,\{1,2,6\},6)^7 (\spaN(5,l_2) \spabtt(l_1,\{1,2,6\},2) \spbN(5,l_1)
   \spbN(6,l_2)-\spaN(5,l_1) \spabtt(l_2,\{1,2,6\},2) \spbN(5,l_2)
   \spbN(6,l_1))}{\spaN(5,l_1) \spaN(5,l_2) \spaN(l_1,l_2) \spabtt(5,\{1,2,6\},2)
   \spabtt(5,\{1,2,6\},6) \spabtt(l_1,\{1,2,6\},2) \spabtt(l_2,\{1,2,6\},2) \spabtt(l_2,\{1,2,6\},6)
   \spbN(1,2) \spbN(1,6)^2 \t(1,2,6)}\!\left[\frac{i \spabtt(l_2,\{1,2,6\},6)}{\spabtt(l_1,\{1,2,6\},6)}\right]^{A},
}\equn\label{6ptmultiplet}
$$
where $A=4-2h$.

\section{Integral Functions}

\subsection{Box Functions}

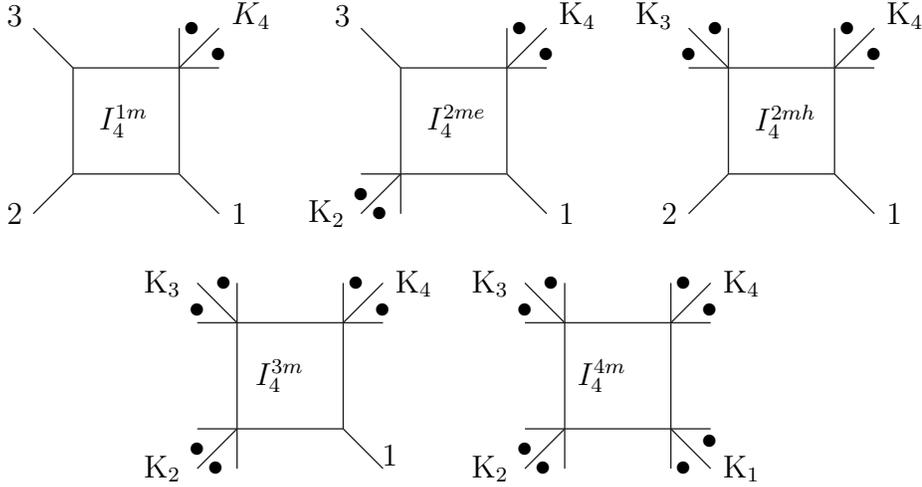
\begin{center}

\begin{picture}(120,95)(0,0)
\Line(30,20)(70,20)
\Line(30,60)(70,60)
\Line(30,20)(30,60)
\Line(70,60)(70,20)

\Line(30,20)(15,5)
\Line(30,60)(15,75)
\Line(70,20)(85,5)
\Line(70,60)(85,75)

\Line(70,60)(70,75)
\Line(70,60)(85,60)
\Text(75,75)[c]{\small$\bullet$}
\Text(85,65)[c]{\small $\bullet$}

\Text(5,5)[l]{\small $\hbox{\rm 2}$}
\Text(90,5)[l]{\small $\hbox{\rm 1}$}
\Text(90,80)[l]{\small ${K_4}$}
\Text(5,80)[l]{\small $\hbox{\rm 3}$}

\Text(40,40)[l]{$I^{1m}_{4}$}
\end{picture}
\begin{picture}(120,95)(0,0)
\Line(30,20)(70,20)
\Line(30,60)(70,60)
\Line(30,20)(30,60)
\Line(70,60)(70,20)

\Line(30,20)(15,5)
\Line(30,60)(15,75)
\Line(70,20)(85,5)
\Line(70,60)(85,75)

\Line(30,20)(30,5)
\Line(30,20)(15,20)
\Text(22,5)[c]{\small$\bullet$}
\Text(15,12)[c]{\small $\bullet$}

\Line(70,60)(70,75)
\Line(70,60)(85,60)
\Text(75,75)[c]{\small$\bullet$}
\Text(85,65)[c]{\small $\bullet$}

\Text(-5,5)[l]{\small ${\rm K_2}$}
\Text(90,5)[l]{\small $\hbox{\rm 1}$}
\Text(5,80)[l]{\small $\hbox{\rm 3}$}
\Text(90,80)[l]{\small ${\rm K_4}$}

\Text(40,40)[l]{$I^{2me}_{4}$}
\end{picture}
\begin{picture}(120,95)(0,0)
\Line(30,20)(70,20)
\Line(30,60)(70,60)
\Line(30,20)(30,60)
\Line(70,60)(70,20)

\Line(30,20)(15,5)
\Line(30,60)(15,75)
\Line(70,20)(85,5)
\Line(70,60)(85,75)

\Line(30,60)(30,75)
\Line(30,60)(15,60)
\Text(15,65)[c]{\small$\bullet$}
\Text(25,75)[c]{\small $\bullet$}

\Line(70,60)(70,75)
\Line(70,60)(85,60)
\Text(75,75)[c]{\small$\bullet$}
\Text(85,65)[c]{\small $\bullet$}

\Text(5,5)[l]{\small $\hbox{\rm 2}$}
\Text(90,5)[l]{\small $\hbox{\rm 1}$}
\Text(90,80)[l]{\small ${\rm K_4}$}
\Text(-5,80)[l]{\small ${\rm K_3}$}

\Text(40,40)[l]{$I^{2mh}_{4}$}
\end{picture}

\begin{picture}(120,95)(0,0)
\Line(30,20)(70,20)
\Line(30,60)(70,60)
\Line(30,20)(30,60)
\Line(70,60)(70,20)

\Line(30,20)(15,5)
\Line(30,60)(15,75)
\Line(70,20)(85,5)
\Line(70,60)(85,75)

\Line(30,20)(30,5)
\Line(30,20)(15,20)
\Text(22,5)[c]{\small $\bullet$}
\Text(15,12)[c]{\small $\bullet$}

\Line(30,60)(30,75)
\Line(30,60)(15,60)
\Text(15,65)[c]{\small $\bullet$}
\Text(25,75)[c]{\small $\bullet$}

\Line(70,60)(70,75)
\Line(70,60)(85,60)
\Text(75,75)[c]{\small $\bullet$}
\Text(85,65)[c]{\small $\bullet$}

\Text(-5,5)[l]{\small  ${\rm K_2}$}
\Text(85,10)[l]{\small ${\rm 1}  $}
\Text(90,75)[l]{\small ${\rm K_4}$}
\Text(-5,75)[l]{\small  ${\rm K_3}$}

\Text(37,40)[l]{$I^{3m}_{4}$}
\end{picture}
\begin{picture}(120,95)(0,0)
\Line(30,20)(70,20)
\Line(30,60)(70,60)
\Line(30,20)(30,60)
\Line(70,60)(70,20)

\Line(30,20)(15,5)
\Line(30,60)(15,75)
\Line(70,20)(85,5)
\Line(70,60)(85,75)

\Line(30,20)(30,5)
\Line(30,20)(15,20)
\Text(22,5)[c]{\small $\bullet$}
\Text(15,12)[c]{\small $\bullet$}

\Line(30,60)(30,75)
\Line(30,60)(15,60)
\Text(15,65)[c]{\small $\bullet$}
\Text(25,75)[c]{\small $\bullet$}

\Line(70,60)(70,75)
\Line(70,60)(85,60)
\Text(75,75)[c]{\small $\bullet$}
\Text(85,65)[c]{\small $\bullet$}

\Line(70,20)(70,5)
\Line(70,20)(85,20)
\Text(75,5)[c]{\small $\bullet$}
\Text(85,15)[c]{\small $\bullet$}

\Text(-5,5)[l]{\small  ${\rm K_2}$}
\Text(90,5)[l]{\small ${\rm K_1}$}
\Text(90,75)[l]{\small ${\rm K_4}$}
\Text(-5,75)[l]{\small  ${\rm K_3}$}

\Text(35,40)[l]{$I^{4m}_{4}$}
\end{picture}
\end{center}

The scalar box integrals considered here have vanishing internal
masses, but may have up to four non-vanishing external masses. Again
by external masses we mean off-shell legs with $K^2 \not =0$. These
integrals are defined and given in~\cite{IntegralsLong} (the
four-mass box was computed by Denner, Nierste, and
Scharf~\cite{FourMassBox}) and are shown in the figures above.

The scalar box integral is,
$$
I_4 = -i \L4\pi\R^{2-\e} \,\int {d^{4-2\e}p\over \L2\pi\R^{4-2\e}}
\;{1\over p^2 \L p-K_1\R^2 \L p-K_1-K_2\R^2 \L p+K_4\R^2}\;.
\equn
$$
The external momentum arguments, $K_{i}$,
are sums of external momenta $k_i$.
In general the integrals are functions of the momentum invariants $K_i^2$ together with
$S\equiv (K_1+K_2)^2$ and $T=(K_2+K_3)^2$.
The no-mass box is, to
${\cal O}(\e^0)$ ,
$$
\eqalign{
I^{0\rm m}_4 [1]\ & =\ \rg \, {1\over s t}
\biggl\{
{2 \over \eps^2} \Bigl[ ( -s)^{-\eps}+ (-t)^{-\eps} \Bigr]
- \ln^2\L {-s \over - t} \R - \pi^2 \biggr\} \ , \cr}
\equn
$$ where $s = (k_1 + k_2)^2$ and $t = (k_2 + k_3)^2$ are the usual
Mandelstam variables. The factor $r_\Gamma$ arises within dimensional
regularisation and is,
$$
r_\Gamma\; = \;
{1\over(4 \pi)^{2\,-\,\eps}}\,{\Gamma(1\,+\,\eps)\,\Gamma^2(1\,-\,\eps)\over\Gamma(1\,-\,2\,\eps)}\,.
\equn
$$
This function appears only in four-point amplitudes with massless particles.

With the labelling of legs shown above, the scalar box integrals,
$I_4$, expanded to ${\cal O}(\e^0)$ for the different cases
reduce to, \
$$\hspace{-1cm}
\eqalign{
  I_{4}^{1{\rm m}} &=\ { -2 \rg  \over  S T }
 \biggl\{
 -{1\over\e^2} \Bigl[ (-S)^{-\e} +
(-T )^{-\e} - (-K_4^2)^{-\e} \Bigr] \cr
 &\ + \Li_2\left(1-{ K_4^2 \over S}\right)
  \ + \ \Li_2\left(1-{K_4^2 \over T }\right)
  \ +{1\over 2} \ln^2\left({ S  \over T }\right)\
+\ {\pi^2\over6} \biggr\} \ ,
\cr}
\equn
$$

$$\hspace{-.75cm}
\eqalign{
 I_{4}^{2{\rm m}e}
&=\
{-2 \rg
     \over S T  -K_2^2 K_4^2 }
\biggl\{
 - {1\over\e^2} \Bigl[ (-S)^{-\e} + (-T)^{-\e}
              - (-K_2^2 )^{-\e} - (-K_4^2 )^{-\e} \Bigr] \cr
&\ +\ \Li_2\left(1-{K_2^2 \over S }\right)
 \ +\ \Li_2\left(1-{K_2^2 \over T}\right)
 \ +\ \Li_2\left(1-{ K_4^2  \over  S }\right)
\cr
&\
 \ +\ \Li_2\left(1-{K_4^2  \over T }\right)
-\ \Li_2\left(1-{ K_2^2 K_4^2
\over  S T }\right)
   \ +\ {1\over 2} \ln^2\left({ S  \over T }\right)
\biggr\} \ ,
\cr }
\equn
$$
$$\hspace{-0.3cm}
\eqalign{ \hskip -1.8 truecm
\!\!  I_{4}^{2{\rm m}h}
&=\ { -2 \rg  \over S T  }
\biggl\{
 -{1\over\e^2} \Bigl[ (- S)^{-\e} + (-T)^{-\e}
              - (-K_3^2 )^{-\e} - (-K_4^2)^{-\e} \Bigr]
\cr &
  \ -\ {1\over2\e^2}
    { (-K_3^2)^{-\e}(-K_4^2)^{-\e}
     \over (- S)^{-\e} }
  \ +\ {1\over 2} \ln^2\left({ S \over T  }\right)
  \ +\ \Li_2\left(1-{ K_3^2 \over  T }\right)
  \ +\ \Li_2\left(1-{ K_4^2 \over T  }\right)
  \biggr\}  \ ,
\cr}
\equn
$$

$$\hspace{-0.18cm}
\eqalign{ \hskip -1.4 truecm
  I_{4}^{3{\rm m}}
&=\ { -2 \rg
     \over S T -K_2^2 K_4^2   }
\biggl\{
 -{1\over\e^2} \Bigl[ (-S )^{-\e} + (-T)^{-\e}
     - (-K_2^2)^{-\e}
     - (-K_3^2 )^{-\e}
     - (-K_4^2 )^{-\e} \Bigr] \cr
\null   & \;\;\;\;\;\;\;\;\;\;
\ -\ {1\over2\e^2}
   { (-K_2^2)^{-\e}(-K_2^2 )^{-\e} \over(-T)^{-\e} }
  \ -\ {1\over2\e^2}
    {(-K_3^2)^{-\e}(-K_4^2)^{-\e}
           \over (-T)^{-\e} }
      \ +\ {1\over2}\ln^2\left({ S \over T }\right)
\cr
  &\;\;\;\;\;\;\;\;\;\;
 +\ \Li_2\left(1-{ K_2^2 \over S  }\right)
   \ +\ \Li_2\left(1-{ K_4^2  \over T }\right)
  \ -\  \Li_2
\left(1-{ K_2^2 K_4^2  \over S T }\right)
\biggr\}\ ,
 \cr}
\equn
$$

$$\hspace{-0.9cm}
\eqalign{
I_{4}^{4{\rm m}} = &
{-\rg  \over  S \; T \;\rho}
\biggl\{ - \Li_2\left(\hf(1-\lambda_1+\lambda_2+\rho)\right)
  \ +\ \Li_2\left(\hf(1-\lambda_1+\lambda_2-\rho)\right) \cr
 &\ -\ \Li_2\left(
   \textstyle-{1\over2\lambda_1}(1-\lambda_1-\lambda_2-\rho)\right)
 \ +\ \Li_2\left(\textstyle-{1\over2\lambda_1}(1-\lambda_1-
    \lambda_2+\rho)\right) \cr
  &\ -\ {1\over2}\ln\left({\lambda_1\over\lambda_2^2}\right)
   \ln\left({ 1+\lambda_1-\lambda_2+\rho \over 1+\lambda_1
        -\lambda_2-\rho }\right) \biggr\} \ ,
   \cr}
\equn
$$
where,
$$
 \rho\ \equiv\ \sqrt{1 - 2\lambda_1 - 2\lambda_2
+ \lambda_1^2 - 2\lambda_1\lambda_2 + \lambda_2^2}\ ,
\equn
$$
and,
$$
\lambda_1 = { K_2^2  \; K_4^2  \over S  \; T  } \; , \hskip 1.5 cm
\lambda_2 = { K_1^2 \;  K_3^2 \over   S \; T  }\ .
\equn
$$

When checking the soft divergences of the seven-point amplitude we
need the  $1/\eps$ singularities
arising from soft singularities in the loop integration.
For the boxes relevant to the seven-point amplitude these are,
$$
\eqalign{
I^{abc\{defg\} } |_{1/\eps}
&
=
-{2 \over s_{ab}s_{bc} (4\pi)^2} \Bigl[
{ \ln(-s_{ab})+\ln(-s_{bc})-\ln(-t_{abc}) \over \eps}
\Bigr]\,,
\cr
I^{a(bc)(def)g} |_{1/\eps}
&
=
-{2 \over s_{ag} t_{abc}  (4\pi)^2} \Bigl[
{ \ln(-s_{ag} )+2\ln(-t_{abc})-\ln(-s_{bc} )-\ln(-t_{def} ) \over 2\eps}
\Bigr]\,,
\cr
I^{a(bc)d(efg)} |_{1/\eps}
&=
-{2 \over (t_{abc}t_{bcd} -s_{bc}t_{efg} )(4\pi)^2 } \Bigl[
{
\ln(-t_{abc}) +\ln(-t_{bcd}) -\ln(-s_{bc}) -\ln(-t_{efg})  \over \eps }
\Bigr],
\cr
I^{a(bc)(de)(fg)} |_{1/\eps}
&
=
-{2 \over( t_{abc}t_{fga} -s_{bc}s_{fg})(4\pi)^2  }
\Bigl[{
\ln(-t_{abc}) +\ln(-t_{fga}) -\ln(-s_{bc}) -\ln(-s_{fg})  \over 2 \eps} )
\Bigr].
\cr}
\equn
$$

\subsection{Triangle and Bubble integral Functions}

Triangle integral functions may have one, two or three massless legs:
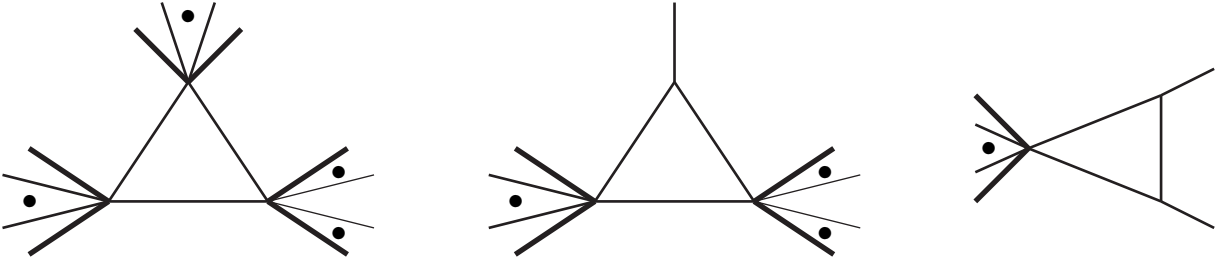
\vspace{-0.3cm}
\begin{center}
\begin{picture}(180,60)(-20,50)
\Line(30,30)(70,40)
\Line(30,30)(70,20)
\SetWidth{2}
\Line(30,30)(60,50)
\Line(30,30)(60,10)
\SetWidth{1}
\Line(30,30)(-30,30)
\Line(-30,30)(0,75)
\Line(30,30)(0,75)
\Line(-30,30)(-70,40)
\Line(-30,30)(-70,20)
\SetWidth{2}
\Line(-30,30)(-60,50)
\Line(-30,30)(-60,10)
\SetWidth{1}
\Text(-60,30)[]{$\bullet$}
\Line(0,75)(-10,105)
\Line(0,75)(10,105)
\SetWidth{2}
\Line(0,75)(-20,95)
\Line(0,75)(20,95)
\Text(0,100)[]{$\bullet$}
\Text(57,41)[]{$\bullet$}
\Text(57,18)[]{$\bullet$}
\end{picture}
\begin{picture}(150,60)(-20,50)
\Line(30,30)(70,40)
\Line(30,30)(70,20)
\SetWidth{2}
\Line(30,30)(60,50)
\Line(30,30)(60,10)
\SetWidth{1}
\Line(30,30)(-30,30)
\Line(-30,30)(0,75)
\Line(30,30)(0,75)
\Line(-30,30)(-70,40)
\Line(-30,30)(-70,20)
\SetWidth{2}
\Line(-30,30)(-60,50)
\Line(-30,30)(-60,10)
\SetWidth{1}
\Text(-60,30)[]{$\bullet$}
\Line(0,75)(0,105)
\SetWidth{2}
\Text(57,41)[]{$\bullet$}
\Text(57,18)[]{$\bullet$}
 \SetWidth{1}
\end{picture}
\begin{picture}(40,60)(-20,50)
\SetWidth{1}
\Line(30,30)(30,70)
\Line(30,70)(-20,50)
\Line(30,30)(-20,50)
\Line(30,30)(50,20)
\Line(30,70)(50,80)
\Line(-20,50)(-40,59)
\Line(-20,50)(-40,41)
\Text(-35,50)[]{$\bullet$}
\SetWidth{2}
\Line(-20,50)(-40,70)
\Line(-20,50)(-40,30)
\end{picture}
\end{center}
\vspace{1.2cm}

The one-mass triangle depends only on the momentum invariant of the
massive leg,
$$
I_{3}^{1\rm m} = {\rg\over\e^2} (-K_1^2)^{-1-\eps} \ .
\equn
$$
The next integral function is
the two-mass triangle integral,
$$
I_{3}^{2 \rm m} = {\rg\over\e^2}
{(-K_1^2)^{-\eps}-(-K_2^2)^{-\eps} \over  (-K_1^2)-(-K_2^2) }\ .
\equn
$$

Note that the one and two mass triangles are linear combinations of
the set of functions,
$$
G(-K^2)= \rg { (-K^2)^{-\eps}  \over\e^2} \; ,
\equn
$$
with,
$$
I^{1m}_3 = G(-K_1^2)\;\; ,
\;\;
I^{2m}_3 ={ 1 \over (-K_1^2)-(-K_2^2) }
\left( G(-K_1^2)-G(-K_2^2  ) \right).
\equn
$$
The $G(-K^2)$ are labelled by the independent momentum invariants
$K^2$ and in fact form an independent basis of functions, unlike the
one and two-mass triangles which are not all independent. For
example, for six-point kinematics there are only twenty-five
independent options for $K^2$ corresponding to 15 independent
$s_{ij}$'s  and  10 independent $t_{ijk}$'s, whereas there are 15
one-mass triangles and 60 two-mass triangles.

The final scalar triangle is the three-mass integral function.
The evaluation of this integral is more
involved,
and can be obtained from ~\cite{ThreeMassTriangle,IntegralsLong},
$$
I_{3}^{3 \rm m}
=\ {i\over \sqrt{\Delta_3}}  \sum_{j=1}^3
  \left[ \Li_2\left(-\left({1+i\delta_j \over 1-i\delta_j}\right)\right)
       - \Li_2\left(-\left({1-i\delta_j \over 1+i\delta_j}\right)\right)
  \right]\ + \ \Ord(\e),
\equn
$$
where,
$$
\eqalign{
\delta_1 & = { K_1^2 - K_2^2 - K_3^2 \over
\sqrt{\Delta_3}} \; , \cr
\delta_2 & = {-K_1^2 + K_2^2 - K_3^2 \over
\sqrt{\Delta_3}} \; , \cr
\delta_3 & = {-K_1^2 - K_2^2 + K_3^2 \over
\sqrt{\Delta_3}}\; ,  \cr}
\equn
$$
and
$$
\Delta_3\equiv
-(K_1^2)^2-(K_2^2)^2-(K_3^2)^2+2 ( K_1^2K_2^2 +K_3^2K_1^2 +K_2^2K_3^2).
\equn
$$

Finally, the bubble integral is,
$$
\eqalign{
I_{2}(K^2) & = {\rg \over \eps (1-2\eps)}(-K^2)^{-\eps}. \cr
}\equn
$$

\end{document}